\documentclass{jfm}

\usepackage{graphicx}
\usepackage{epstopdf, epsfig}
\usepackage{amsmath} % for diamonds
\usepackage{xcolor}
\usepackage{bm} % for the \boldsymbol

\usepackage{xcolor}

\newcommand{\RomanNumeralCaps}[1]
\linenumbers

\usepackage{tikz}
\usetikzlibrary{positioning,shapes}
\usepackage{soul}

\usepackage[font=small,labelfont=bf,
   justification=justified,
   format=plain]{caption}
   
\title{Energy transfer mechanisms in adverse pressure gradient turbulent boundary layers  }

\author{Taygun R. Gungor \aff{1,2},
 Yvan Maciel\aff{2}, \and
 Ayse G. Gungor\aff{1}  \corresp{\email{ayse.gungor@itu.edu.tr}}}

\affiliation{\aff{1} Faculty of Aeronautics and Astronautics, Istanbul Technical University, 34469 Maslak, Istanbul, Turkey
\aff{2}Department of Mechanical Engineering, Laval University, Quebec City, QC, G1V 0A6 Canada}

\shorttitle{Energy transfer mechanisms in APG TBLs}	
\shortauthor{T. R. Gungor, Y. Maciel, and A. G. Gungor}

\newcommand{\mi}{\boldsymbol{-} \mathrel{\mkern -14mu} \boldsymbol{-}}

\definecolor{col1}{rgb}{          0    0.4470    0.7410}
\definecolor{col2}{rgb}{  0.8500    0.3250    0.0980}
\definecolor{col3}{rgb}{    0.9290    0.6940    0.1250 }
\definecolor{col4}{rgb}{      0.4940    0.1840    0.5560}
\definecolor{col5}{rgb}{      0.4660    0.6740    0.1880}
\definecolor{col6}{rgb}{      0.3010    0.7450    0.9330}
\definecolor{col7}{rgb}{      0.6350    0.0780    0.1840}

\begin{document}

\maketitle

\begin{abstract}

The energy transfer mechanisms and structures playing a role in these mechanisms in adverse-pressure-gradient (APG) turbulent boundary layers (TBLs) with small and large velocity defects are investigated. We examine the wall-normal and spectral distributions of energy, production and pressure-strain in APG TBLs and compare these distributions with those in canonical flows. It is found that the spectral distributions of production and pressure-strain are not profoundly affected by an increase of the velocity defect, although the energy spectra drastically change in the inner layer of the large defect APG TBL. In the latter, the signature of the inner layer streaks is absent from the energy spectra. However, the production and pressure-strain spectra suggest that the near-wall cycle or another energy transfer mechanism with similar spectral features still exist in large defect TBLs. In the outer layer, energetic, production and pressure-strain structures appear to change from wall-attached to wall-detached structures with increasing velocity defect. Despite this, the 2D spectral distributions have similar shapes and wavelength aspect ratios of the peaks in all these flows. These observations suggest that outer layer energy transfer mechanisms may be the same in wall-bounded flows, no matter if dynamically relevant structures are attached or detached to the wall. Therefore, the conclusion is that the mechanisms responsible for turbulence production and inter-component energy transfer may remain the same within each layer in all these flows. It is the intensity of these mechanisms within one layer that changes with velocity defect, because of the local mean shear variation.

\end{abstract}


%%%%%%%%%%%%%%%%%%%%%%%%%%%%%%%%%%%%%%%%%%%%
%%%%%%%%%%%%%%%%%%%%%%%%%%%%%%%%%%%%%%%%%%%%
%%%%%%%%%%%%%%%%%%%%%%%%%%%%%%%%%%%%%%%%%%%%
%%%%%%%%%%%%%%%%%%%%%%%%%%%%%%%%%%%%%%%%%%%%
%%%%%%%%%%%%%%%%%%%%%%%%%%%%%%%%%%%%%%%%%%%%
%%%%%%%%%%%%%%%%%%%%%%%%%%%%%%%%%%%%%%%%%%%%
%%%%%%%%%%%%%%%%%%%%%%%%%%%%%%%%%%%%%%%%%%%%
%%%%%%%%%%%%%%%%%%%%%%%%%%%%%%%%%%%%%%%%%%%%
%%%%%%%%%%%%%%%%%%%%%%%%%%%%%%%%%%%%%%%%%%%%
%%%%%%%%%%%%%%%%%%%%%%%%%%%%%%%%%%%%%%%%%%%%
%%%%%%%%%%%%%%%%%%%%%%%%%%%%%%%%%%%%%%%%%%%%
%%%%%%%%%%%%%%%%%%%%%%%%%%%%%%%%%%%%%%%%%%%%
%%%%%%%%%%%%%%%%%%%%%%%%%%%%%%%%%%%%%%%%%%%%
%%%%%%%%%%%%%%%%%%%%%%%%%%%%%%%%%%%%%%%%%%%%
%%%%%%%%%%%%%%%%%%%%%%%%%%%%%%%%%%%%%%%%%%%%
%%%%%%%%%%%%%%%%%%%%%%%%%%%%%%%%%%%%%%%%%%%%
%%%%%%%%%%%%%%%%%%%%%%%%%%%%%%%%%%%%%%%%%%%%
%%%%%%%%%%%%%%%%%%%%%%%%%%%%%%%%%%%%%%%%%%%%

\section{Introduction}

When a turbulent boundary layer is subjected to a sufficiently strong or prolonged adverse pressure gradient, its mean momentum defect increases. This growing defect progressively changes the nature of the flow, and hence APG TBLs become different from canonical wall-bounded flows such as zero-pressure-gradient (ZPG) TBLs or channel flows.

Indeed, the change in the mean velocity profile leads to a different distribution of the mean shear in the wall-normal direction, which in turn affects the turbulent energy distribution in the boundary layer. With less mean shear, the turbulent activity in the inner layer decreases \citep{skaare1994turbulent, elsberry2000experimental}, and the inner maximum of Reynolds stresses vanishes in the case of large velocity defect TBLs \citep{maciel2006study,gungor2016scaling, maciel2018outer}. The outer layer turbulence, on the other hand, becomes dominant when the velocity defect is important, and a turbulence production peak emerges in the outer layer \citep{skaare1994turbulent, gungor2016scaling}. 

In the inner layer, the impact of the pressure gradient on the coherent structures also depends on the extent of the mean velocity defect. The spanwise and streamwise sizes of the most energetic $\langle u^2\rangle$-carrying structures do not change significantly when the defect is small \citep{lee2017large,harun2013pressure}. Like in canonical flows, the inner peak of the energy spectra of $\langle u^2\rangle$ is at inner-scaled spanwise wavelength ($\lambda_z^+$) of approximately $120$ \citep{lee2017large,tanarro2020effect} and inner-scaled streamwise wavelength ($\lambda_x^+$) of approximately $1000$ \citep{harun2013pressure,vila2020separating}. Furthermore, the shape of the $\langle u^2\rangle$ spectra is resemblant in ZPG and small defect APG TBLs \citep{harun2013pressure,tanarro2020effect}. When the defect is large, spatial organization and spectral features of the energetic structures change in the inner layer. The near-wall streaks are weakened \citep{skote2002direct, lee2009structure}. They also become more irregular, disorganized, less streaky \citep{lee2009structure,maciel2017structural}, and even vanish at separation \citep{rahgozar2012statistical}. Moreover, the inner peak in the $\langle u^2\rangle$ spectra, which is connected to the streaks, vanishes in the large-defect case \citep{lee2017large,kitsios2017direct}. Apart from the changes in the streaks, \cite{maciel2017structural, maciel2017coherent} reported that sweeps and ejections are weaker in the inner layer than in the outer layer in large-defect TBLs, and their number is less than in ZPG TBLs.

The elevated outer layer turbulence activity in APG TBLs was reported numerous times regardless of the velocity defect \citep{harun2013pressure, lee2017large, tanarro2020effect}. The $\langle u^2\rangle$ spectra show that the energetic outer layer structures are longer and wider than their counterparts in the inner layer in APG TBLs \citep{harun2013pressure,kitsios2017direct}. The outer peak of the spectra is at $\lambda_x/\delta\approx3$ \citep{harun2013pressure,vila2020separating} and $\lambda_z/\delta\approx1$ \citep{bobke2017history,lee2017large}, where $\delta$ is the boundary layer thickness. Elevated outer layer activity is also reported for high Reynolds number canonical flows \citep{hutchins2007large,marusic2010high} with an outer peak in the $\langle u^2\rangle$ spectra due to elongated, meandering structures \citep{marusic2010high}, which are called superstructures in external flows \citep{hutchins2007evidence} and very large-scale motions (VLSMs) in internal flows \citep{kim1999very}. The streamwise wavelength of the outer peak in the energy spectra of $\langle u^2\rangle$, which is associated with these structures, is in the order of $6\delta$ in boundary layers \citep{smits2011high}. There is also another type of structures, which are called large scale motions (LSMs), associated with smaller wavelengths of $\lambda_x/\delta\approx 2$-$3$ \citep{smits2011high}. The energetic outer layer structures in APG TBLs are more similar to LSMs than superstructures/VLSMs in terms of dimensions. However, \cite{vila2020separating} investigated the effects of APG and high Reynolds number for small-defect APG TBLs and found two outer peaks (one similar to the ZPG superstructures' one, and the typical APG one) in the $\langle u^2\rangle$ spectra at sufficiently high Reynolds number. Regarding the magnitude of the typical APG outer peak, several studies have reported its increase with increasing velocity defect \citep{lee2017large,vila2020experimental} when the levels are normalized with friction-viscous scales. However, such a conclusion could be an artefact of using the friction-viscous scales because they are not proper scales for energy levels for APG TBLs \citep{gungor2016scaling}, especially for large defect ones \citep{maciel2018outer}.

The turbulence regeneration mechanisms, or in other words, self-sustaining mechanisms, have been extensively studied in canonical flows. However, studies concerning APG TBLs are rather scarce. One of the mechanisms that has been suggested for APG TBLs is the instability of streaks \citep{marquillie2011instability}, which was already considered as a regeneration mechanism for canonical wall-bounded flows \citep{hall1991linear}. Another mechanism proposed for the outer region of large defect APG TBLs is an inflectional instability associated with the outermost inflection point of the mean velocity profile, which is inviscidly unstable. \cite{elsberry2000experimental} suggested that such an instability affects the flow field in their large defect APG TBL. The location of the inflection point was close to that of the maxima of the Reynolds normal stresses and turbulent kinetic energy production. The presence of an inflection point at the same wall-normal location as the peak of the Reynolds stresses was reported by other researchers as well \citep{gungor2016scaling,kitsios2017direct}. Furthermore, \cite{schatzman2017experimental} suggested that an embedded shear layer, which is centered around the aforementioned inflection point in the outer layer, exists in APG TBLs. They proposed scaling parameters based on this idea of an embedded shear layer and obtained self-similar mean velocity and Reynolds stress profiles in their large-defect APG TBL, as well as in other large-defect cases. \cite{balantrapu2021structure} also obtained self-similar profiles with the same scaling in their highly decelerated axisymmetric turbulent boundary layer. However, \cite{maciel2017coherent} noted that they could not find any roller-like structures that are the sign of a Kelvin-Helmholtz type instability. In addition, it is important to note that moderate-defect APG TBLs have an outer maximum of the Reynolds stresses without the presence of an inviscidly unstable inflection point in the mean velocity profile \citep{maciel2018outer}. It is still not known whether an inflection point instability exists, and furthermore if it is directly responsible for turbulent regeneration in large defect APG TBLs, but it is possible that the presence of Reynolds stress peaks and an inflection point in the outer layer is simply correlated without any causality \citep{balantrapu2021structure}.

From a different perspective that does not necessarily involve an inflectional instability, several researchers have reported that APG TBLs with large velocity defect might behave like free shear flows due to the change in the mean velocity \citep{gungor2016scaling,kitsios2017direct}. \cite{gungor2020reynolds} demonstrated that the Reynolds shear stress carrying structures in large defect APG TBLs and homogeneous shear turbulence flow, which has no inflection point, have similar shapes and are mostly dependent on the local mean shear. These studies suggest that self-sustaining mechanisms might be similar in the outer layer of large defect APG TBLs and free shear flows and highlight the causal role played by mean shear.

The task of identifying the self-sustaining mechanisms present in APG TBLs is a formidable one. A first step along that route is to better understand the energy transfers directly resulting from the self-sustaining processes, namely turbulence production and energy transfer between Reynolds stress components (the pressure-strain term in the Reynolds stress transport equations). Examining the wall-normal distributions of production and pressure-strain does not suffice for that purpose. A more complete picture can be obtained by also investigating the spectral distributions, among scales and position, of these energy transfers. The idea of studying the spectra of the terms in the Reynolds stress transport equations was introduced first by \cite{lumley1964spectral}, and it has gained attention recently. The spectra of the turbulent kinetic energy or Reynolds stress transport equations have been utilized to examine spatial transport, energy cascade, or scale separation in channel flows \citep{mizuno2016spectra,cho2018scale,lee2019spectral}, Couette flows \citep{kawata2018inverse} or ZPG TBLs \citep{chan2021interscale}. As mentioned above, in the present work we focus on the spectral features of production and pressure-strain with the aim of better understanding the self-sustaining mechanisms present in APG TBLs.

More precisely, the goal of this study is to examine the spectral distributions of energy, production and pressure-strain in inner and outer layers of APG TBLs and compare them with the ones in canonical wall-bounded flows to better understand the mechanisms involved. We consider both small and large velocity defect cases to analyze the effect of velocity defect on these energy transfer mechanisms. 

\color{black}

%%%%%%%%%%%%%%%%%%%%%%%%%%%%%%%%%%%%%%%%%%%%
%%%%%%%%%%%%%%%%%%%%%%%%%%%%%%%%%%%%%%%%%%%%
%%%%%%%%%%%%%%%%%%%%%%%%%%%%%%%%%%%%%%%%%%%%
%%%%%%%%%%%%%%%%%%%%%%%%%%%%%%%%%%%%%%%%%%%%
%%%%%%%%%%%%%%%%%%%%%%%%%%%%%%%%%%%%%%%%%%%%
%%%%%%%%%%%%%%%%%%%%%%%%%%%%%%%%%%%%%%%%%%%%
%%%%%%%%%%%%%%%%%%%%%%%%%%%%%%%%%%%%%%%%%%%%
%%%%%%%%%%%%%%%%%%%%%%%%%%%%%%%%%%%%%%%%%%%%
%%%%%%%%%%%%%%%%%%%%%%%%%%%%%%%%%%%%%%%%%%%%
%%%%%%%%%%%%%%%%%%%%%%%%%%%%%%%%%%%%%%%%%%%%
%%%%%%%%%%%%%%%%%%%%%%%%%%%%%%%%%%%%%%%%%%%%
%%%%%%%%%%%%%%%%%%%%%%%%%%%%%%%%%%%%%%%%%%%%
%%%%%%%%%%%%%%%%%%%%%%%%%%%%%%%%%%%%%%%%%%%%
%%%%%%%%%%%%%%%%%%%%%%%%%%%%%%%%%%%%%%%%%%%%
%%%%%%%%%%%%%%%%%%%%%%%%%%%%%%%%%%%%%%%%%%%%
%%%%%%%%%%%%%%%%%%%%%%%%%%%%%%%%%%%%%%%%%%%%
%%%%%%%%%%%%%%%%%%%%%%%%%%%%%%%%%%%%%%%%%%%%
%%%%%%%%%%%%%%%%%%%%%%%%%%%%%%%%%%%%%%%%%%%%

% \clearpage

\section{Databases}\label{sec:rules_submission}

We utilized three types of flows in this paper: APG TBLs, ZPG TBLs, and channel flows. A new DNS database is generated for APG TBLs. This new database is an extension of the database that was introduced by \cite{gungor2017direct} and analyzed by \cite{maciel2018outer}, which was referred to as {DNS2017}. The main difference between the new database and DNS2017 is that the new database has a wider spanwise width than the previous one. Other than that, both databases are almost identical with minor differences in the boundary conditions. Since they are almost identical, we will only use here the most recent one. The details of the databases employed in this paper are given in the following sections.

\begin{figure}
       \begin{tikzpicture}   
              \centering                     
\node(a){ \includegraphics[scale=0.6]{ 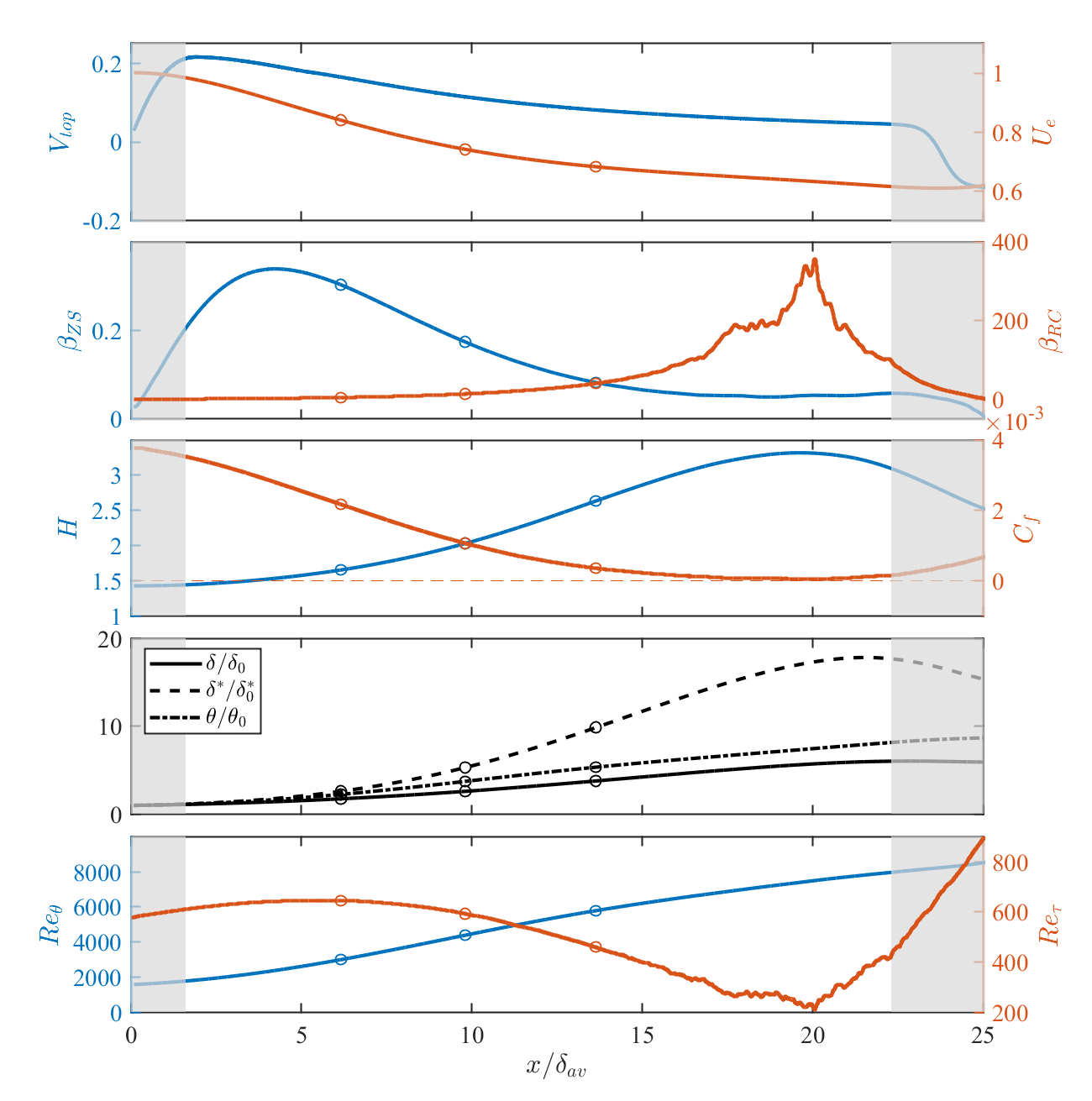}};
\node at  (-6.5,  6) [overlay, remember picture] {$(a)$};
\node at  (-6.5,  3.6) [overlay, remember picture] {$(b)$};
\node at  (-6.5,  1.15) [overlay, remember picture] {$(c)$};
\node at  (-6.5,  -1.32) [overlay, remember picture] {$(d)$};
\node at  (-6.5,  -3.6) [overlay, remember picture] {$(e)$};
       \end{tikzpicture}  
\caption{Streamwise development of the main parameters of the APG TBL. The suction/blowing boundary condition ($V_{top}$) and edge velocity ($U_e$) $(a)$. Pressure gradient parameters $\beta_{ZS}$ and $\beta_{RC}$ $(b)$. Shape factor, $H$, and skin friction coefficient, $C_f$ $(c)$. $\delta$, $\delta^*$ and $\theta$ normalized with their value at inlet $(d)$. Reynolds numbers $\Rey_\theta$ and $\Rey_\tau$. The white zone shows the useful range. The open circles denote the three streamwise positions that are used in the paper. }
\label{fig:chars}
\end{figure}
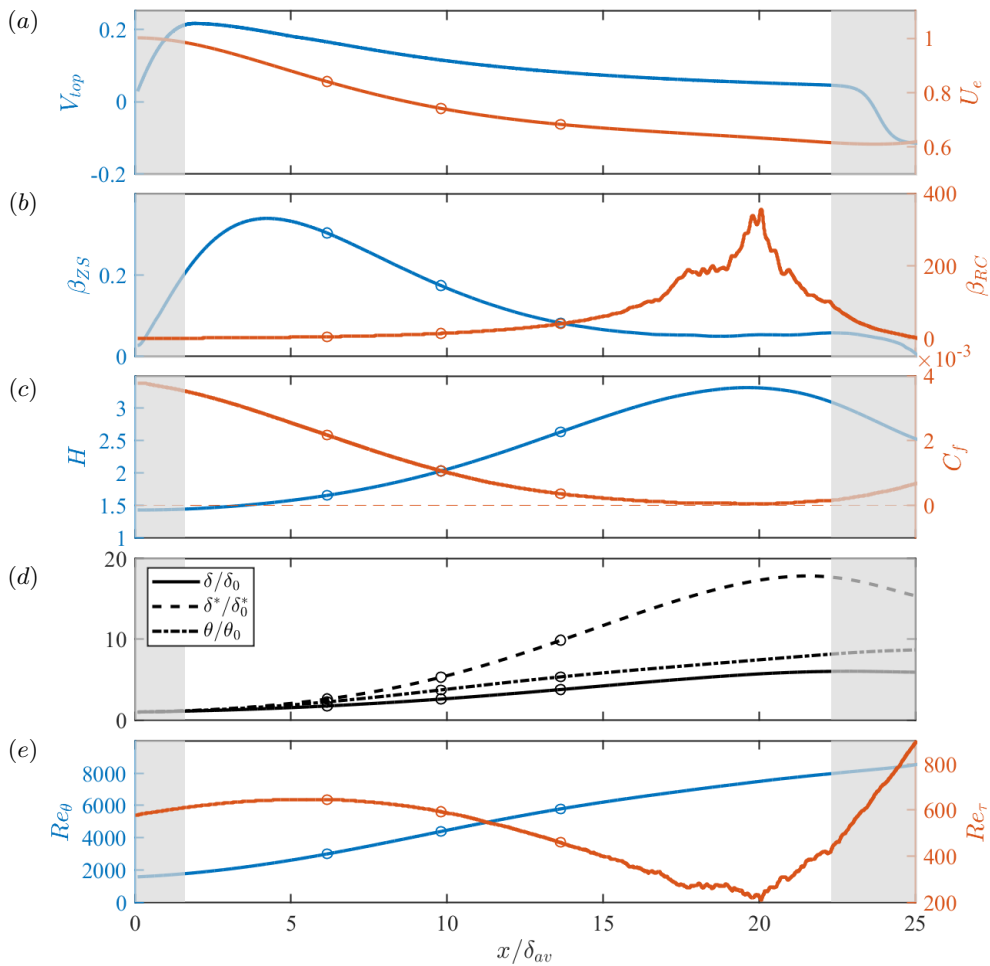

\subsection{The current APG TBL}

The current database is a DNS database of a non-equilibrium APG TBL with a Reynolds number based on momentum thickness ($\Rey_\theta$) reaching up to $8000$. The code used to perform the DNS solves the three-dimensional incompressible Navier-Stokes equations. It provides the time evolution of three-dimensional velocity and pressure fields for a given flow configuration. The DNS code employs a hybrid MPI/OpenMP approach for parallelization. Further details on the code can be found in \cite{simens2009high} and \cite{borrell2013code}.

The DNS computational setup consists of two simulation domains, the auxiliary and main domains, running concurrently as described in \cite{sillero2013one} and \cite{gungor2016scaling}. The auxiliary ZPG TBL DNS, with a coarser resolution, is intended to provide the realistic turbulent inflow data for the main APG TBL DNS. The goal of using two domains is to provide inflow conditions for the main DNS with lower computational cost. Regarding the other boundary conditions, the bottom surface is a flat plate with a no-slip boundary condition. The side boundary conditions are periodic. The far-field boundary condition is adjusted in the form of suction and blowing to apply adverse/favorable pressure gradients. The outflow condition is a Neumann boundary condition. The computational domain of the main APG TBL simulation, a rectangular volume with a streamwise, wall-normal and spanwise length of $(L_x$, $L_y$, $L_z$)/$\delta_{av}$ = $25.3$, $4.8$, $7.1$, is discretized with $N_x$, $N_y$, $N_z$ = $4609\times736\times1920$ grid points. Here the average boundary layer thickness ($\delta_{av}$) is calculated within the useful range (or so-called domain of interest), which is shown as a white zone in figure \ref{fig:chars}. \color{black}

Figure \ref{fig:chars} shows the spatial development of main parameters of the APG TBL. As stated before, the pressure gradient in the APG TBL is generated by imposing a suction/blowing boundary condition for the wall-normal component of the velocity at the top boundary ($V_{top}$).  Figure \ref{fig:chars}a shows the spatial development of $V_{top}$ along with the edge velocity, $U_e$. The definition of $U_e$ is not straightforward in flows with inviscid velocity that varies in the wall-normal direction, such as the current flow. Although it is not the most generally applicable definition (see for instance \cite{griffin2021general}), we utilize the maximum streamwise component of the mean velocity in the wall-normal direction as $U_e$ for this study to be consistent with most literature \citep{gungor2016scaling,maciel2018outer} and because it works for the flow cases studied here. \color{black}

 At the beginning of the domain, the suction velocity sharply increases within a short distance to impose the desired pressure gradient distribution. Downstream of this region, the suction velocity is adjusted to obtain a regularly increasing shape factor $H$ (figure \ref{fig:chars}c). The change in $V_{top}$ near the end of the computational domain is to increase the numerical stability of the simulation by accelerating the flow so that the Neumann boundary condition still holds. The suction boundary condition decelerates the flow, which is seen from the development of the edge velocity ($U_e$) in figure \ref{fig:chars}a.

Figure \ref{fig:chars}b shows two pressure gradient parameters that characterize the effect of the pressure gradient on the outer layer: Rotta-Clauser pressure gradient parameter ($\beta_{RC}$) and the pressure gradient parameter based on the Zagarola-Smits velocity ($U_{ZS}$). They are defined in equation \ref{pres}. 

\begin{equation}
\beta_{RC}= \frac{\delta^*}{\rho u_\tau^2}\frac{dP_e}{dx}, \hspace{0.5cm} 
\beta_{ZS}= \frac{\delta}{\rho U_{ZS}^2}\frac{dP_e}{dx}
\label{pres}
\end{equation}

\noindent Here, $\delta^*$ is the displacement thickness, $\rho$ is the density, $u_\tau$ is the friction velocity and $P_e$ is the pressure at the edge of boundary layer. $\beta_{RC}$, which is the traditionally used pressure gradient parameter, progressively increases until $\beta_{RC}$ is above $350$ and then sharply decreases. However, as \cite{maciel2018outer} demonstrated, it is not a valid pressure gradient parameter for TBLs with large velocity defect. $\beta_{ZS}$, which correctly represents the local impact of the pressure gradient on the boundary layer regardless of velocity defect \citep{maciel2018outer}, increases near the flow entrance up to 0.35 and then decreases to approximately 0.05 and remains mostly the same in the last part of the domain. The rapid increase of $\beta_{ZS}$ at the beginning of the domain causes the increase in momentum loss in the boundary layer ($H$ increases, figure \ref{fig:chars}c). The subsequent decrease of $\beta_{ZS}$ over most of the domain should theoretically lead to momentum gain in the boundary layer (H decreases). However, because the boundary layer responds with a delay to the pressure force evolution, the momentum gain only occurs at the end of the domain.

The effect of the pressure gradient on the flow is also seen in the development of the skin friction coefficient, $C_f$, as shown in figure \ref{fig:chars}c. $C_f$ steadily decreases until around $x/\delta_{av}\approx20$ and becomes very close to zero. The behavior of $C_f$ and $H$ reflects the strong non-equilibrium nature of the current APG TBL flow.

Figure \ref{fig:chars}d shows the spatial development of $\delta$, $\delta^*$, and momentum ($\theta$) thicknesses. They all increase until close to the end of the domain. Figure \ref{fig:chars}e shows the distributions of the most commonly used Reynolds numbers for TBLs, $\Rey_\theta$ and $\Rey_\tau$. $\Rey_\theta$ increases approximately from 2000 to 8000. Reynods number based on $u_\tau$, $\Rey_\tau$ develops irregularly in the streamwise direction. However, it is not valid because $u_\tau$ is not a valid scale for AGP TBLs with large velocity defect. The irregularity of $\Rey_\tau$ stems from utilizing $u_\tau$ as a velocity scale.

\subsection{Existing Databases}

Because we focus on spectral analysis in the current paper, we have two criteria for selecting existing databases of ZPG TBLs and channel flows: sufficiently high Reynolds number and availability of spectral distributions of Reynolds stresses, production and pressure-strain. The DNS database of \cite{lee2015direct} with a $\Rey_\tau$ of 2000 is employed for channel flows. This database is chosen because some one- and two-dimensional spectra are available as a function of streamwise and spanwise wavenumbers ($k_x$, $k_z$) and $y$. The DNS database of \cite{sillero2013one} with a $\Rey_\theta$ of $6500$ and the experimental database of \cite{baidya2021spanwise} with a $\Rey_\theta$ of $6191$ are chosen for ZPG TBLs. We chose these two databases for ZPG TBLs because the Reynolds numbers are similar and the 1D spectral distributions of Reynolds stress and production are available as a function of $k_z$ and $y$ for the database of \cite{sillero2013one} and $k_x$ and $y$ for the database of \cite{baidya2021spanwise}. They complement each other for the spectral analysis without introducing any significant Reynolds number effect. \color{black}

\begin{table}
  \begin{center}
\def~{\hphantom{0}}
\begin{tabular*}{\textwidth}{ @{\extracolsep{\fill}} llccccccccc}
& Name  &  Type  & Database &  $\Rey_\theta$ & $\Rey_\tau$ &   H & $C_f\times10^{-3}$ & $\beta_{ZS}$ & $\beta_{RC}$ & Color \& Symbols \\[5pt]
&APG1&APG TBL & DNS&   $3005$ & $646$  & 1.65 &  2.1719 & 0.30 & 4.5& \color{col1}\tikz\draw [thick] (0,0) -- (0.5,0);  \\   
&APG2&APG TBL & DNS&   $4395$ & $593$  & 2.00 & 1.0674  & 0.17 &  13.6& \color{col1}\tikz\draw [thick,dashed] (0,0) -- (0.5,0);\\
&APG3&APG TBL & DNS&   $5787$ & $460$  & 2.63 & 0.3599 & 0.08 & 40.1& \color{col1}\tikz\draw [thick,dash dot] (0,0) -- (0.5,0);\\
&ZPGa&  ZPG TBL & DNS& $6500$ & $1990$ & 1.35 & 2.7063 & $\approx 0$ &$\approx 0$ & \color{col2}$\mi$\\
&ZPGb&  ZPG TBL & Exp. &$6191$ &$2493$ & 1.35 &   2.7602 & $\approx 0$ &$\approx 0$ & \color{col2}$\circ$ \\
&CH&  Channel & DNS&NA  &  $1995$ &  NA   &  3.3605 & NA &  NA & \color{col5}\tikz\draw [thick] (0,0) -- (0.5,0); \\
  \end{tabular*}
  \caption{The information about the databases used in the paper.}
  \label{tab22}
  \end{center}
\end{table}

\subsection{Flow Description} 

We aim to investigate a wide range of velocity defect cases in this study. For the APG TBL, three streamwise positions from the APG TBL are employed. These streamwise positions, as shown in figure \ref{fig:chars}, represent three velocity defect cases, and their corresponding shape factors are $1.65$, $2.00$ and $2.63$. The shape factor is $1.35$ for both ZPG TBL cases since they are at similar Reynolds numbers. The velocity defect is smaller in the ZPG TBL than the small defect case of the APG TBL. As for the channel flow, the velocity defect, which is with respect to the centerline velocity, is smaller than those of the ZPG TBLs as it is well-known. These five flow cases, details of which are provided in Table \ref{tab22}, cover a wide range of velocity defect situations from a channel flow to a TBL with a large velocity defect.

Regarding the Reynolds number of the cases, the ZPG TBL and the channel flow cases have similar $\Rey_\tau$ and are selected at higher $\Rey_\tau$ and $\Rey_\theta$ than the APG TBL cases. We could have chosen different databases for ZPG TBLs or channel flows with a lower Reynolds number instead. However, it is better to analyze canonical flows at high Reynolds number because of the elevated outer layer activity in such flows \citep{hutchins2007evidence}. In this manner, we can compare the outer peaks of the spectral distributions of canonical flows and APG TBLs.

Throughout the paper, the streamwise, wall-normal and spanwise directions are referred to as $x$, $y$, and $z$ or $1$, $2$, and $3$ for index notation. The corresponding instantaneous velocity components are $\bar{u}$, $\bar{v}$, and $\bar{w}$. The brackets, $\langle.\rangle$, denote ensemble averaging. Furthermore upper cases and lower cases denote the mean value and the fluctuations, respectively. Thus, $\langle \bar{u}_i\rangle=U_i$ and $\bar{u}_i=U_i+u_i$. The upper index plus means the friction-viscous scales, with the friction velocity $u_\tau$ as the velocity scale and $\nu/u_\tau$ as the length scale.

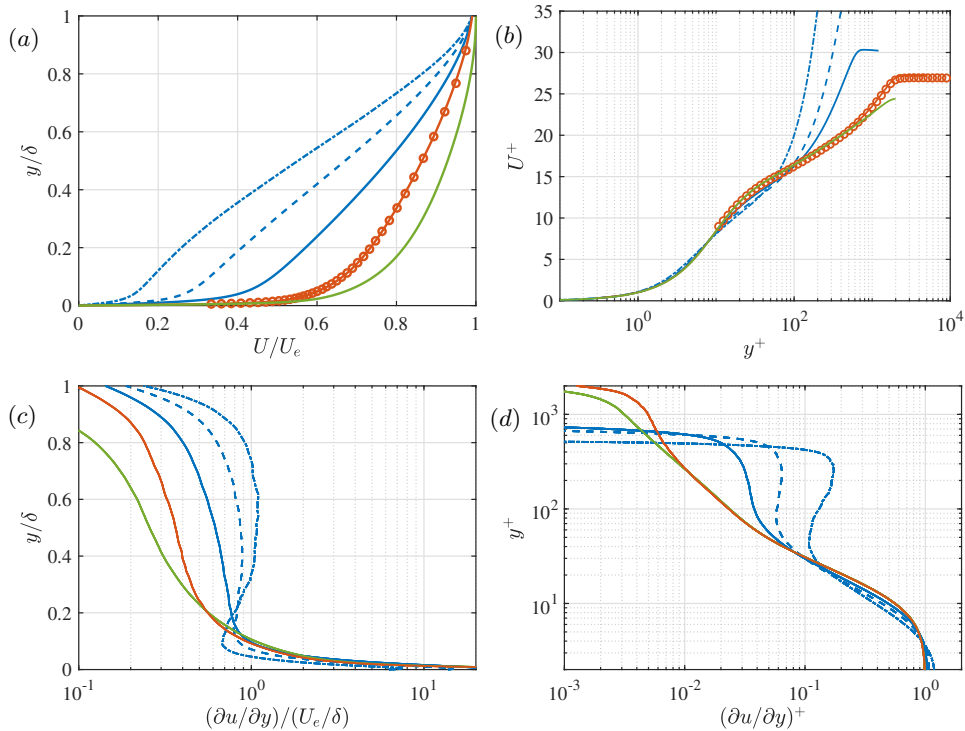
\begin{figure}

       \begin{tikzpicture}   
       \centering                     
\node(a){ \includegraphics[scale=0.45]{ 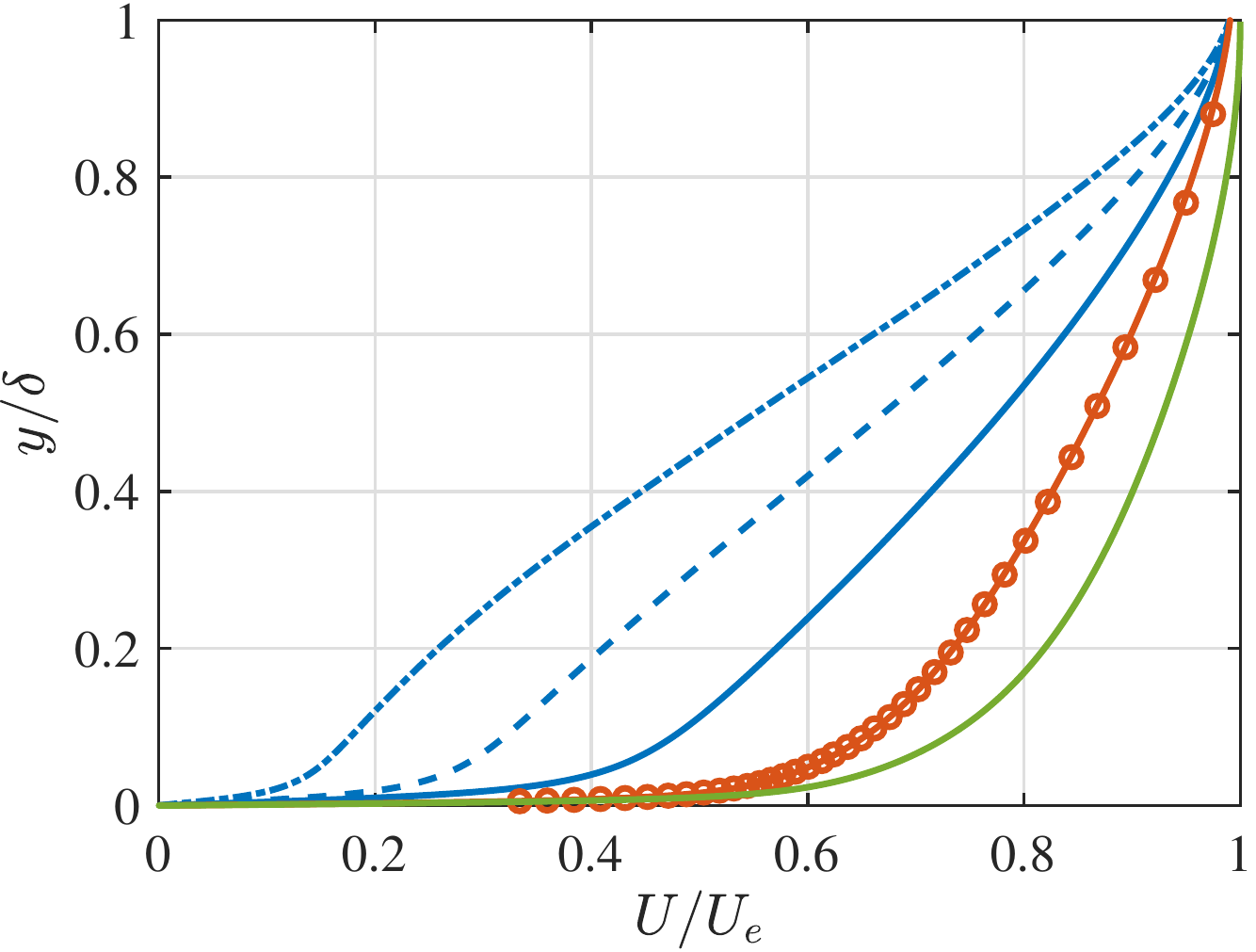}};
\node at  (-3.,  1.9) [overlay, remember picture] {$(a)$};
       \end{tikzpicture}   
                     \begin{tikzpicture}   
       \centering                     
\node(a){ \includegraphics[scale=0.45]{ 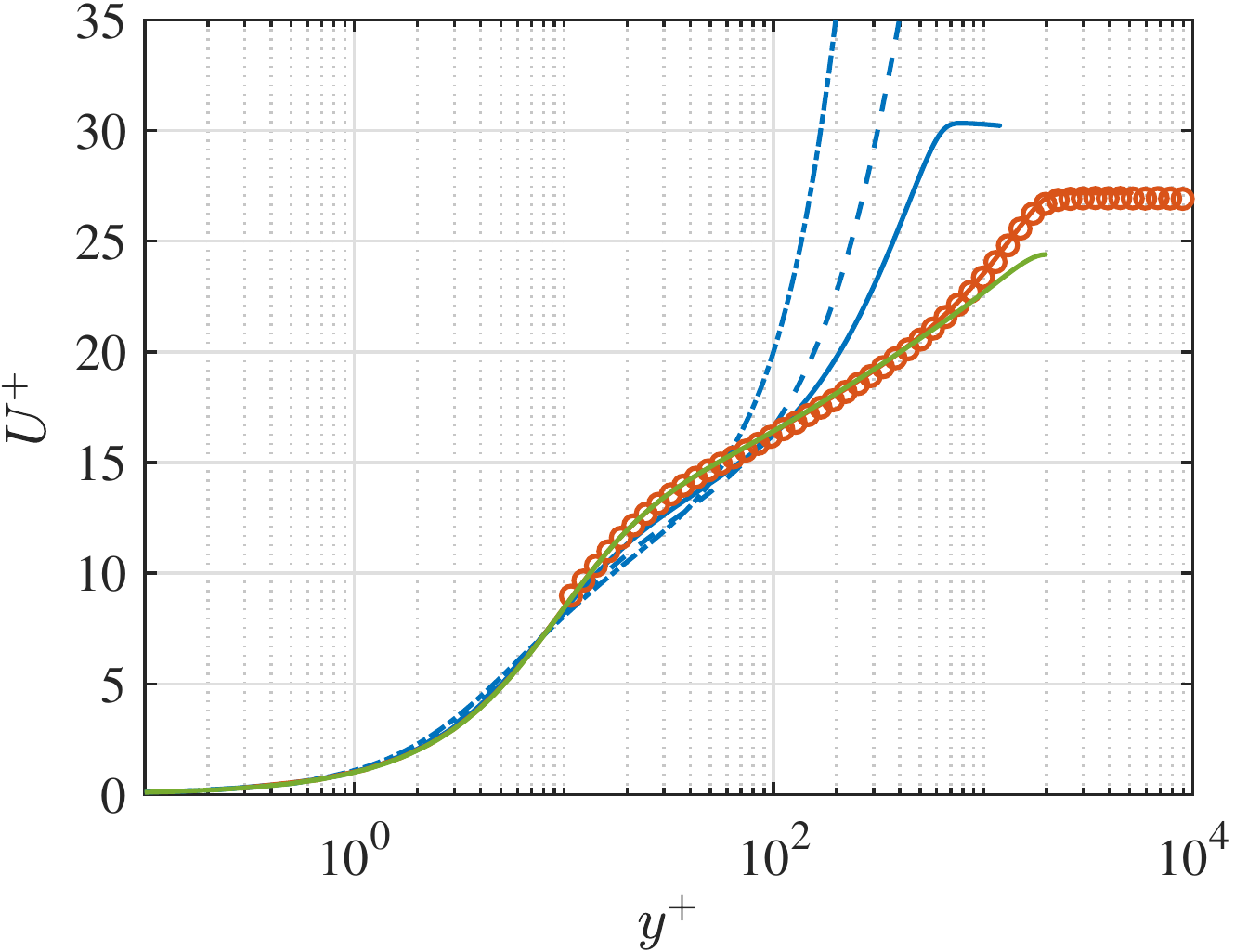}};
\node at  (-3.,  1.9) [overlay, remember picture] {$(b)$};
       \end{tikzpicture} 
       
              \begin{tikzpicture}   
       \centering                     
\node(a){ \includegraphics[scale=0.45]{ 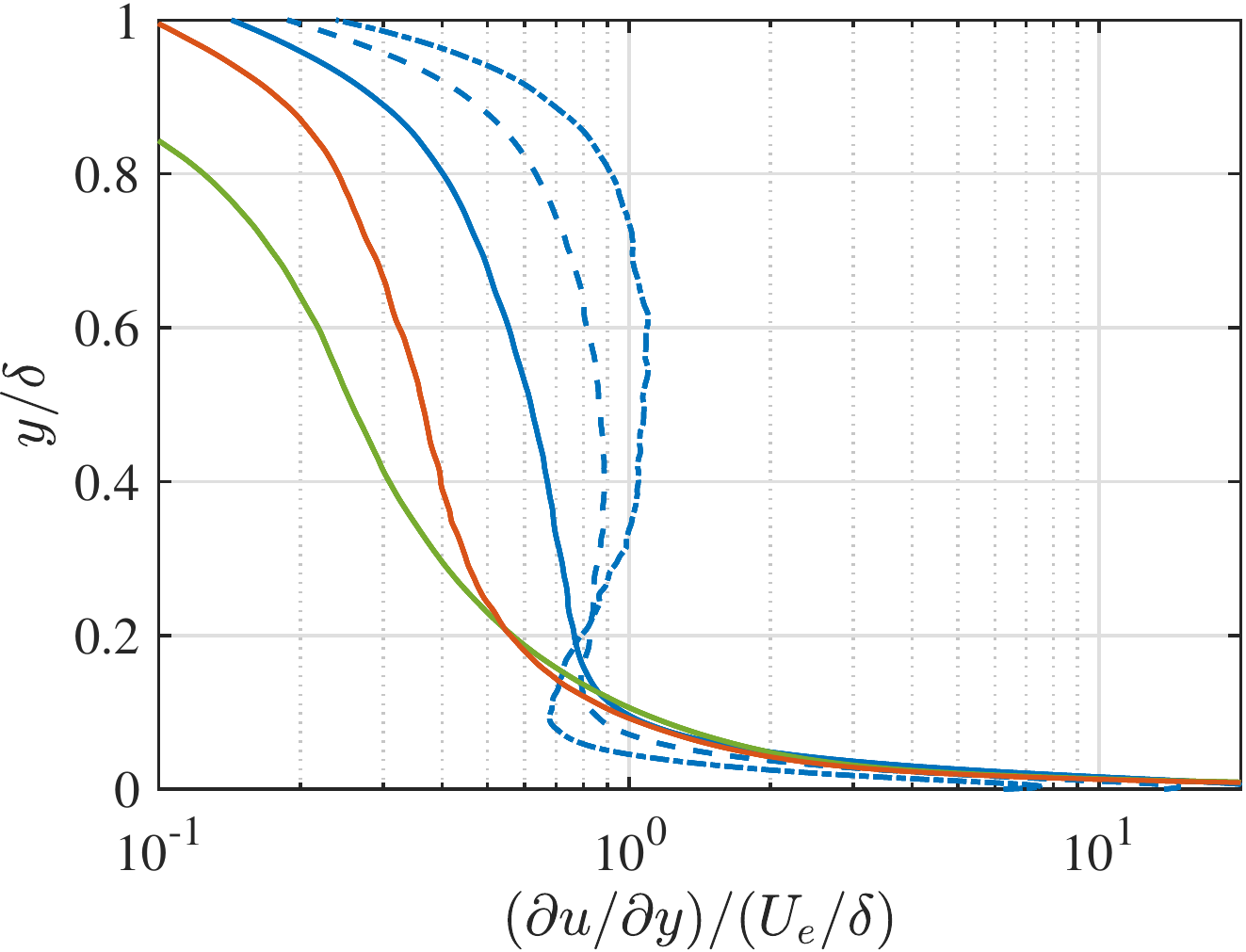}};
\node at  (-3,  1.8) [overlay, remember picture] {$(c)$};
       \end{tikzpicture} 
                     \begin{tikzpicture}   
       \centering                     
\node(a){ \includegraphics[scale=0.45]{ 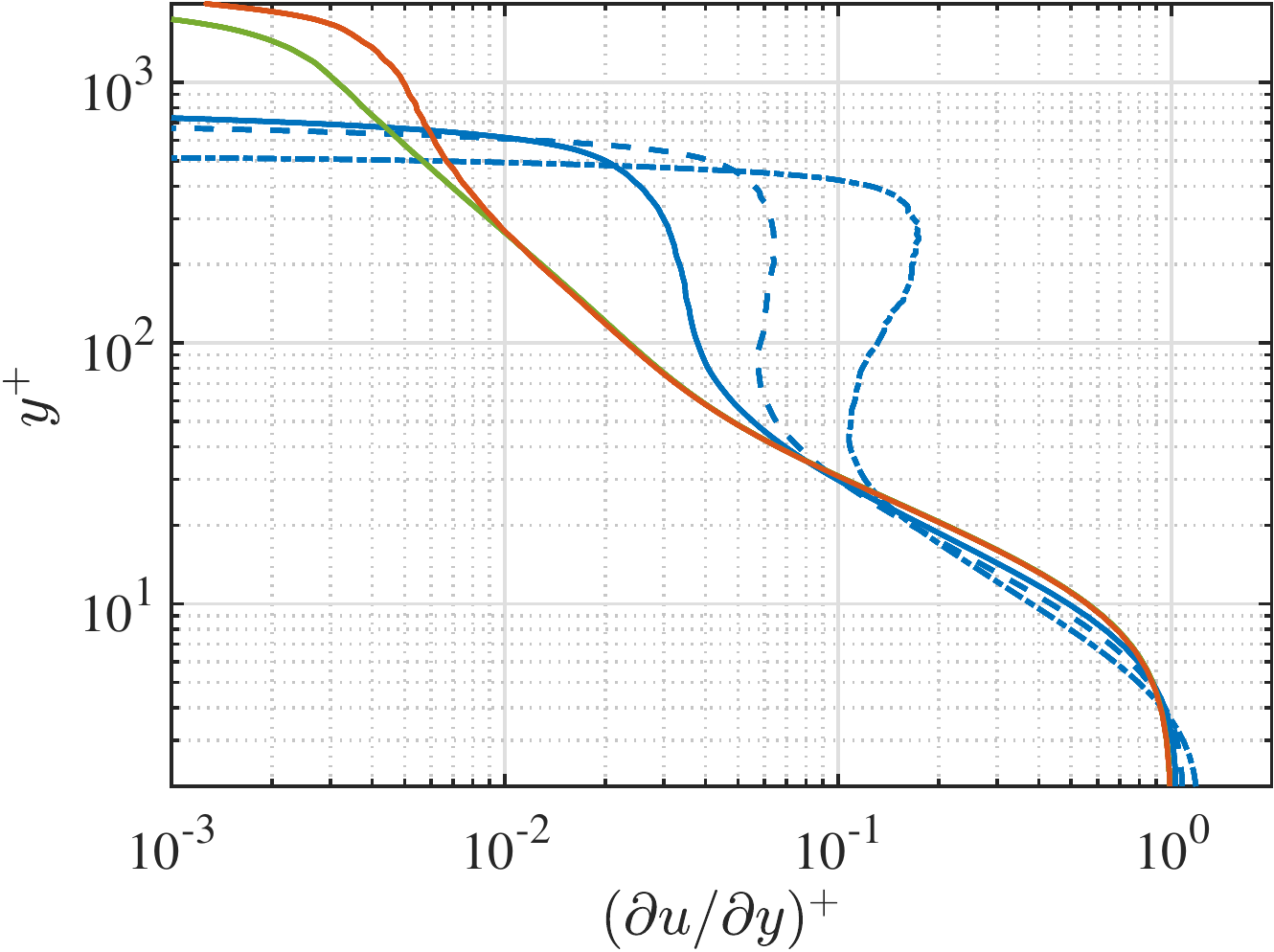}};
\node at  (-3,  1.8) [overlay, remember picture] {$(d)$};
       \end{tikzpicture} 
      
\caption{The mean velocity profiles of all databases normalized with outer scales (a) and friction-viscous units (b). The mean shear profiles of the DNS databases normalized with outer scales (c) and friction-viscous units (d). Legend as in Table 1.}
\label{fig:mv}
\end{figure}

%%%%%%%%%%%%%%%%%%%%%%%%%%%%%%%%%%%%%%%%%%%%
%%%%%%%%%%%%%%%%%%%%%%%%%%%%%%%%%%%%%%%%%%%%
%%%%%%%%%%%%%%%%%%%%%%%%%%%%%%%%%%%%%%%%%%%%
%%%%%%%%%%%%%%%%%%%%%%%%%%%%%%%%%%%%%%%%%%%%
%%%%%%%%%%%%%%%%%%%%%%%%%%%%%%%%%%%%%%%%%%%%
%%%%%%%%%%%%%%%%%%%%%%%%%%%%%%%%%%%%%%%%%%%%
%%%%%%%%%%%%%%%%%%%%%%%%%%%%%%%%%%%%%%%%%%%%
%%%%%%%%%%%%%%%%%%%%%%%%%%%%%%%%%%%%%%%%%%%%
%%%%%%%%%%%%%%%%%%%%%%%%%%%%%%%%%%%%%%%%%%%%
%%%%%%%%%%%%%%%%%%%%%%%%%%%%%%%%%%%%%%%%%%%%
%%%%%%%%%%%%%%%%%%%%%%%%%%%%%%%%%%%%%%%%%%%%
%%%%%%%%%%%%%%%%%%%%%%%%%%%%%%%%%%%%%%%%%%%%
%%%%%%%%%%%%%%%%%%%%%%%%%%%%%%%%%%%%%%%%%%%%
%%%%%%%%%%%%%%%%%%%%%%%%%%%%%%%%%%%%%%%%%%%%
%%%%%%%%%%%%%%%%%%%%%%%%%%%%%%%%%%%%%%%%%%%%
%%%%%%%%%%%%%%%%%%%%%%%%%%%%%%%%%%%%%%%%%%%%
%%%%%%%%%%%%%%%%%%%%%%%%%%%%%%%%%%%%%%%%%%%%
%%%%%%%%%%%%%%%%%%%%%%%%%%%%%%%%%%%%%%%%%%%%

\section{Wall-normal distributions of mean flow and Reynolds stress properties}

In this section, we will first investigate the wall-normal distributions of the mean flow, Reynolds stresses and the Reynolds stress budgets. For all the figures presented here, profiles are plotted using a logarithmic axis for $y^+$ and a linear one for $y/\delta$ to examine inner and outer layers in more detail. Furthermore, the parameters in the inner-scaled profiles are normalized using friction-viscous scales. Although friction-viscous scales are not appropriate scales for the large defect case, they are still employed because there are no suitable inner scales for comparing the small and large defect cases \citep{maciel2018outer}. The edge velocity ($U_e$) and $\delta$ are employed to scale the parameters in the outer region. \color{black}

\subsection{Mean flow}

Figures \ref{fig:mv}a and \ref{fig:mv}b present the outer- and inner-scaled mean velocities as a function of $y$ for all the databases. The outer-scaled mean velocity profiles demonstrate the momentum deficit in the APG TBL cases. As the flow develops under the effect of the APG, the defect in the mean velocity profile increases. The profile starts to resemble velocity profiles of mixing layers \citep{gungor2016scaling} in the large defect case with an inflection point in the middle of the boundary layer. The inner-scaled mean velocity profiles show that the mean velocity deviates from the log law in the APG TBL cases. Moreover, this deviation increases with increasing velocity defect. Furthermore, friction-viscous scales progressively fail to scale the mean velocity in the inner region as the defect increases \citep{gungor2016scaling, maciel2018outer}.

Figures \ref{fig:mv}c and \ref{fig:mv}d show the mean shear profiles for the DNS databases. The mean shear profile for the experimental ZPG TBL case is not given here due to lack of points near the wall.  The mean shear distribution is important because it directly plays a role in turbulence production and hence turbulence in the flow. The change in the mean velocity profiles significantly affects the mean shear distribution in the outer layer. As the defect increases, the mean shear increases in the outer layer. More importantly, the relative magnitude of mean shear in the outer layer with respect to the inner layer increases with increasing velocity defect, as can be seen in figure \ref{fig:mv}d. Regarding the inner layer, the mean shear remains fairly similar when it is normalized with friction-viscous scales as expected, despite the direct effect of the pressure force near the wall in the APG TBL cases. The impact of the varying mean shear on turbulence will be further discussed in section 5.

\begin{figure}
\hspace*{-0.5cm}
       \begin{tikzpicture}   
       \centering                     
\node(a){ \includegraphics[scale=0.7]{ 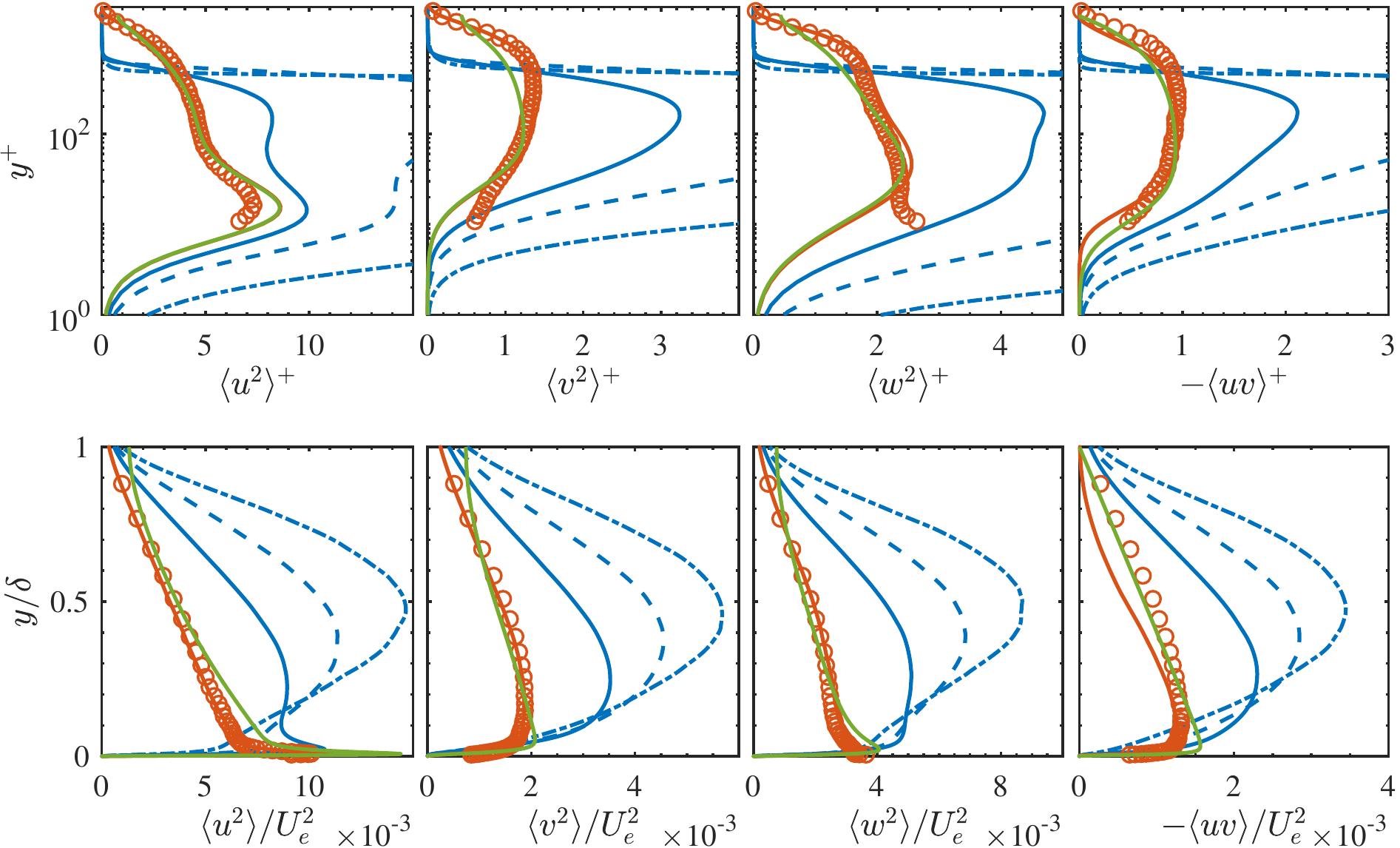}};
\node at  (-6.3,  3.8) [overlay, remember picture] {$(a)$};
\node at  (-6.3,  -0.6) [overlay, remember picture] {$(b)$};
\end{tikzpicture}
\caption{Reynolds stress profiles normalized with friction-viscous units (a) and outer scales (b). Legend as in Table 1.}
\label{fig:rs}
\end{figure}

\subsection{Reynolds stresses}

Figure \ref{fig:rs} presents the wall-normal distribution of the components of the Reynolds-stress tensor for all cases. The inner scales scale the Reynolds stresses well for the canonical flows. The $\langle u^2 \rangle$ profiles of the channel flow and ZPGa collapse perfectly with friction-viscous units, as expected. The $\langle u^2 \rangle$ levels are slightly smaller in ZPGb than in ZPGa in the inner layer, but it is due to the lack of spatial resolution of the probe used in the experiments \citep{baidya2017distance}. 

The change in the mean shear in the APG TBL (figures \ref{fig:mv}c and \ref{fig:mv}d) progressively changes the Reynolds stress profiles. The $\langle u^2 \rangle$ profiles are still fairly similar for APG1 and canonical wall-bounded flows in the inner layer, even if the scaled amplitude increases. The inner peak for $\langle u^2\rangle$ still exists in APG1. However, such a similarity is not encountered for the other components of the Reynolds stress tensor. Moreover, figure \ref{fig:rs}b shows that the turbulent activity in the inner layer diminishes with respect to that in the outer layer as the defect increases. The outer layer becomes dominant as the mean shear increases in the outer layer. In the large defect cases, APG2 and APG3, all components peak in the middle of the boundary layer, where the mean shear has a plateau. Regarding the magnitude of the Reynolds stresses, the levels progressively increase with increasing velocity defect when Reynolds stresses are normalized with friction-viscous scales because they are not appropriate scales for large velocity defect cases \citep{maciel2018outer}. It is important to state that $U_e$ is not necessarily an appropriate outer scale either \citep{maciel2006self}; however, it conserves the order of magnitude of Reynolds stresses in the range of velocity defects and Reynolds numbers of the cases examined here.

The Reynolds stress profiles of the current APG TBL case are consistent with APG TBL cases in the literature. In the small defect case, there is an inner peak for $\langle u^2 \rangle$ and an elevated outer layer activity for all Reynolds stress components. This has been reported for equilibrium \citep{skaare1994turbulent,lee2017large} and non-equilibrium cases \citep{gungor2016scaling} in small defect APG TBLs. Moreover, the $y$-position and energy levels of the inner peak matches well when the velocity defect of the cases are similar \citep{kitsios2016direct,tanarro2020effect}. Regarding the large defect case, other researchers have already reported the increasing importance of the outer layer as the mean shear increases in the outer layer \citep{skaare1994turbulent,gungor2016scaling,kitsios2017direct}.

\subsection{Reynolds stress budgets}

To understand the energy transfer mechanisms in APG TBLs, the budget of the Reynolds stress tensor is investigated first through the transport equations for the Reynolds stresses (equations \ref{ters}).

\begin{equation}\label{ters}
\begin{split}
 	0=&- 
\bigg(\langle u_i u_k\rangle\frac{\partial U_j}{\partial x_k}+ \langle u_j    u_k\rangle\frac{\partial U_i }{\partial x_k} \bigg )   - 
			  \frac{\partial \langle u_iu_ju_k\rangle}{\partial x_k} \\
			  & +	\big \langle \frac{p}{\rho} (\frac{\partial u_i}{\partial x_j}+\frac{\partial u_j}{\partial x_i}) \big \rangle - \frac{1}{\rho} \frac{\partial}{\partial x_k} (\langle u_i p\rangle \delta_{jk}+\langle u_j p\rangle \delta_{ik})
			  \\ & - 	 2\nu \langle \frac{\partial u_i}{\partial x_k} \frac{\partial u_j}{\partial x_k}\rangle+  \nu \nabla^2 \langle u_iu_j\rangle - U_k\frac{\partial \langle u_i u_j \rangle }{ \partial x_k} 
		\end{split}
\end{equation}

\noindent Here $\nu$ is viscosity and $\delta_{ik}$ is the Kronecker delta function. The terms are in order, production, turbulent transport, pressure-strain, pressure transport, viscous dissipation, viscous transport and mean convection.

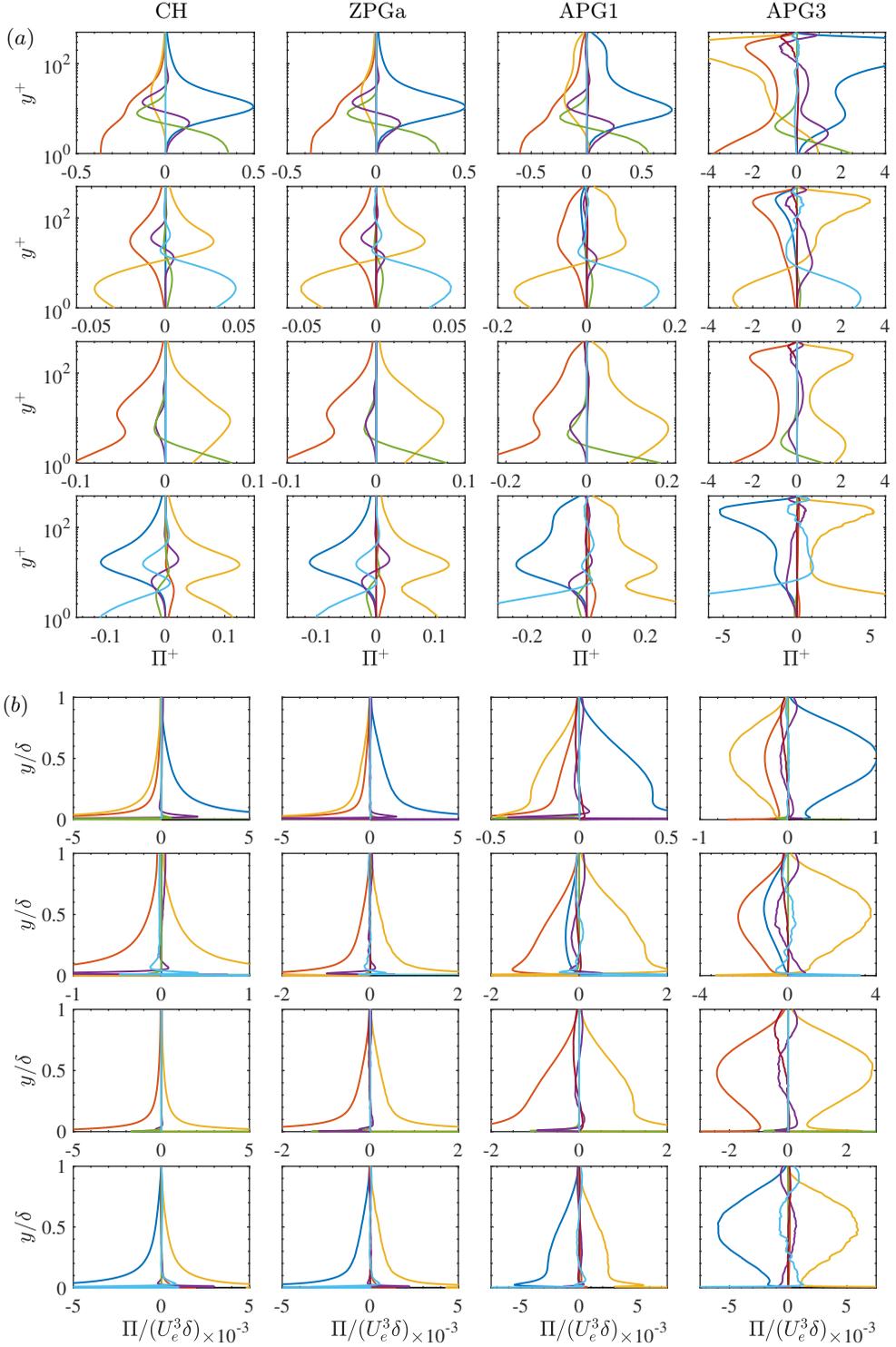
\begin{figure}

\vspace{0.3cm}
       \begin{tikzpicture}   
       \centering                     
\node(a){ \includegraphics[scale=0.5]{ 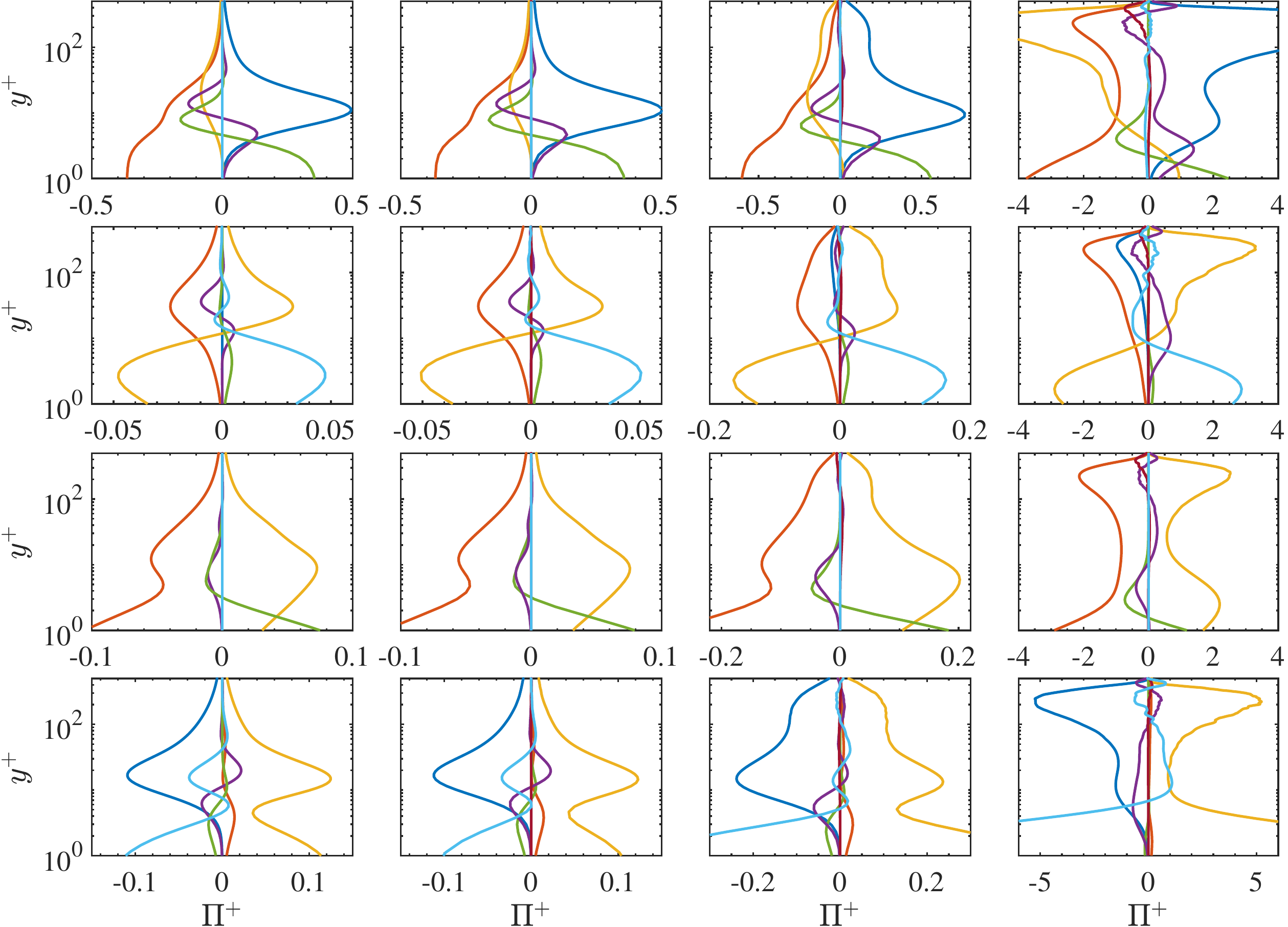}};
\node at  (-4.1,  4.9) [overlay, remember picture] {CH};
\node at  (-1.1,  4.9) [overlay, remember picture] {ZPGa};
\node at  (1.98,  4.9) [overlay, remember picture] {APG1};
\node at  (5.,  4.9) [overlay, remember picture] {APG3};
\node at  (-6.3,  4.5) [overlay, remember picture] {$(a)$};
\end{tikzpicture}

\vspace{0.1cm}
       \begin{tikzpicture}   
       \centering                     
\node(a){ \includegraphics[scale=0.5]{ 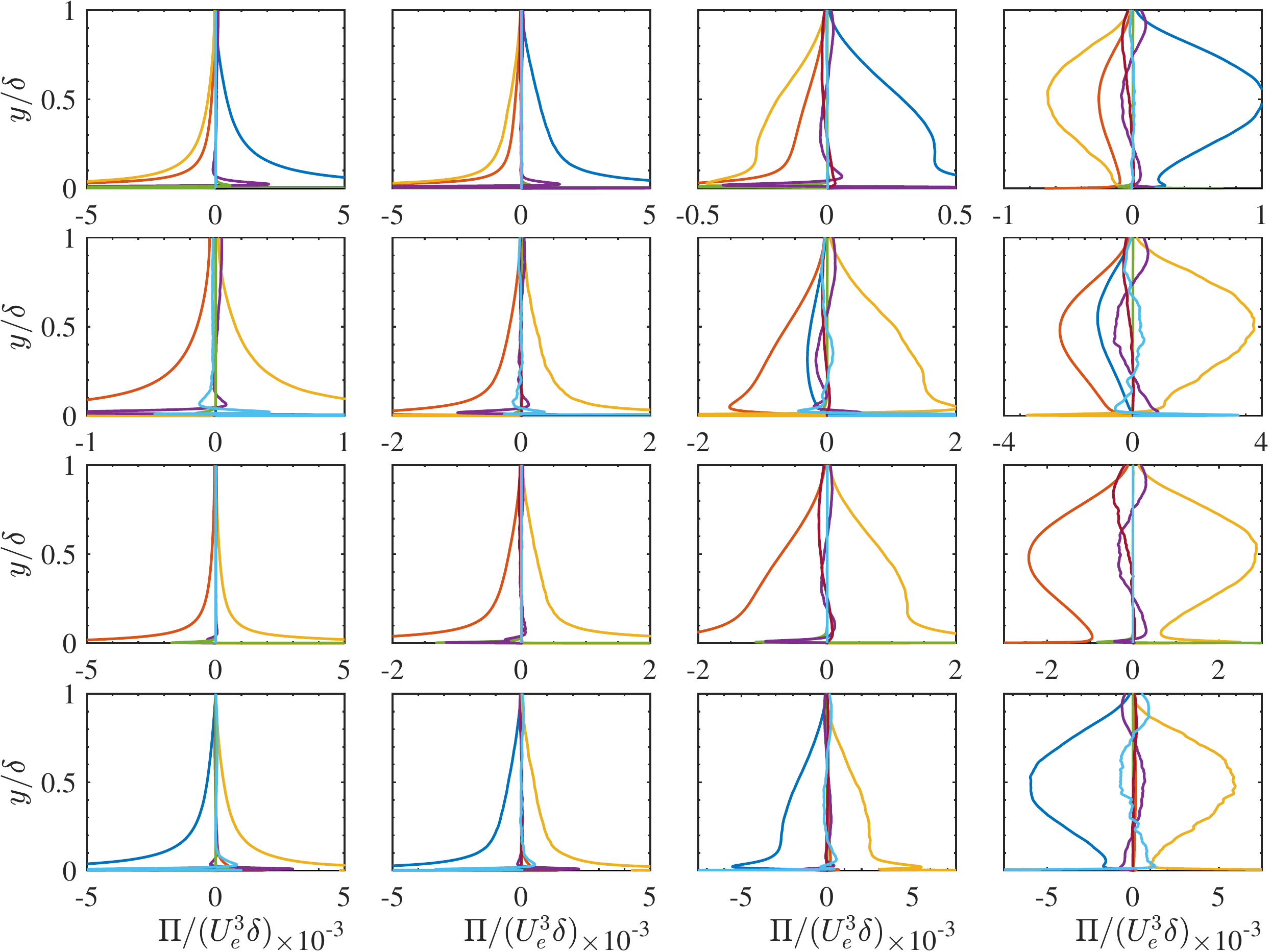}};
\node at  (-6.3,  4.5) [overlay, remember picture] {$(b)$};
\end{tikzpicture}

\caption{The Reynolds stress budgets. The levels and axes are normalized with the friction-viscous scales ($a$) and outer scales ($b$). The rows are for $\langle u^2\rangle$, $\langle v^2\rangle$, $\langle w^2\rangle$, and $\langle uv\rangle$ for each figure. 
Production:\color{col1}$\boldsymbol{-}$\color{black}, 
Dissipation:\color{col2}$\boldsymbol{-}$\color{black}, 
Pressure-strain:\color{col3}$\boldsymbol{-}$\color{black}, 
Viscous diffusion:\color{col5}$\boldsymbol{-}$\color{black}, 
Turbulent transport:\color{col4}$\boldsymbol{-}$\color{black},
Mean convection:\color{col7}$\boldsymbol{-}$\color{black}, 
Pressure transport:\color{col6}$\boldsymbol{-}$\color{black}.\color{black} }
\label{fig:bud_outer}
\end{figure}

Figure \ref{fig:bud_outer} presents Reynolds stress budgets for canonical flows, and small- and large-defect APG TBLs. The profiles are plotted with logarithmic and linear axes to emphasize inner and outer layer, as mentioned before. The Reynolds stress budget distributions show that energy follows the same main path in both inner and outer layers, regardless of the flow as expected. Energy is extracted from the mean flow through $\langle u^2 \rangle$ production. Some of this energy is transferred to $\langle v^2 \rangle$ and $\langle w^2 \rangle$ through pressure strain, which is a sink term for $\langle u^2 \rangle$ and a source term for the other normal components except in the very near-wall region. Energy is also transported or dissipated in all components. Furthermore, $\langle v^2 \rangle$ production, which is zero for channel flows and negligible for ZPG TBLs, is a sink term that transfers energy back to the mean flow in the outer region for APG TBLs. Regarding the Reynolds shear stress, the production and pressure strain are almost in balance in all cases except in the near-wall region.

The magnitude of the budget terms are very similar for CH and ZPGa for the inner-scaled profiles. However, the levels drastically increase as the defect increases when normalized with $u_\tau$, consistently with the trend we observe in the Reynolds stress profiles, which again confirms that $u_\tau$ is not a proper inner scale for APG TBLs with large velocity defect.

In the inner region, as figure \ref{fig:bud_outer}a shows, the behaviour of the Reynolds stress budgets of the canonical flows and APG1 are almost the same for all normal components regardless of the velocity defect. The shape of the budget profiles is almost identical, with a minor shift in $y^+$. For instance, the $\langle u^2 \rangle$ production peak is at $y^+\approx 11$ and $9$ in canonical flows and APG1, respectively. A clear difference between canonical flows and APG1 is that the turbulent and pressure transport above $y^+\approx 10$ for $\langle v^2 \rangle$ is negligible for APG1.

As the mean shear distribution changes from APG1 to APG3, the overall behaviour of the source and sink terms (production, pressure strain and dissipation) in the inner layer does not considerably change except for the aforementioned difference in magnitude. Besides the magnitude change, the main difference is that they start to increase with $y$ above $y^+=30$ because they all peak in the outer layer, as is discussed later. The change of relative importance between the inner and outer layer affects turbulent transport, as well. It becomes a gain term for $\langle u^2 \rangle$ and $\langle v^2 \rangle$ in most of the inner layer due to the elevated turbulent activity in the outer layer of the large defect case. Energy is transported from the outer layer to the inner layer in APG3, unlike what happens in the other cases.

Regarding the outer layer, as shown in the outer-scaled profiles in figure \ref{fig:bud_outer}b, production, pressure strain and dissipation are at high levels between $y/\delta=0.1$ and $0.3$ in APG1, much like the Reynolds stresses themselves, but no peak is present in the outer layer. Despite the lack of an outer peak in APG1, a plateau between approximately $y/\delta=0.1$ and $0.3$ that is not observed in the channel flow and the ZPG TBL exists in production and pressure-strain terms. In the large defect case, an outer peak emerges for the source and sink terms at around $y/\delta=0.5$. Such a behaviour has been reported before \citep{skaare1994turbulent,gungor2016scaling,kitsios2017direct}.

%%%%%%%%%%%%%%%%%%%%%%%%%%%%%%%%%%%%%%%%%%%%
%%%%%%%%%%%%%%%%%%%%%%%%%%%%%%%%%%%%%%%%%%%%
%%%%%%%%%%%%%%%%%%%%%%%%%%%%%%%%%%%%%%%%%%%%
%%%%%%%%%%%%%%%%%%%%%%%%%%%%%%%%%%%%%%%%%%%%
%%%%%%%%%%%%%%%%%%%%%%%%%%%%%%%%%%%%%%%%%%%%
%%%%%%%%%%%%%%%%%%%%%%%%%%%%%%%%%%%%%%%%%%%%
%%%%%%%%%%%%%%%%%%%%%%%%%%%%%%%%%%%%%%%%%%%%
%%%%%%%%%%%%%%%%%%%%%%%%%%%%%%%%%%%%%%%%%%%%
%%%%%%%%%%%%%%%%%%%%%%%%%%%%%%%%%%%%%%%%%%%%
%%%%%%%%%%%%%%%%%%%%%%%%%%%%%%%%%%%%%%%%%%%%
%%%%%%%%%%%%%%%%%%%%%%%%%%%%%%%%%%%%%%%%%%%%
%%%%%%%%%%%%%%%%%%%%%%%%%%%%%%%%%%%%%%%%%%%%
%%%%%%%%%%%%%%%%%%%%%%%%%%%%%%%%%%%%%%%%%%%%
%%%%%%%%%%%%%%%%%%%%%%%%%%%%%%%%%%%%%%%%%%%%
%%%%%%%%%%%%%%%%%%%%%%%%%%%%%%%%%%%%%%%%%%%%
%%%%%%%%%%%%%%%%%%%%%%%%%%%%%%%%%%%%%%%%%%%%
%%%%%%%%%%%%%%%%%%%%%%%%%%%%%%%%%%%%%%%%%%%%
%%%%%%%%%%%%%%%%%%%%%%%%%%%%%%%%%%%%%%%%%%%%

\section{Spectral Analysis }

The Reynolds stress and Reynolds stress budget profiles provide information about the wall-normal distributions of energy and energy transfer, however not about the coherent structures that carry energy or play a role in these energy transfer mechanisms. To investigate those coherent structures, the spectral distributions of production and pressure-strain of the Reynolds stress tensor are analyzed using the transport equation for the two-point velocity correlation tensor, along with the spectral distribution of energy. The reason for using the two-point correlation equation is that the spectral information is directly linked to the two-point correlations. 

\color{black}

\subsection{Methodology}

We will first present the transport equation for the two-point correlation tensor\color{black}. Let $x_i$ and $\tilde{x_i}$ be the components of the coordinates of the two points used to compute the correlations. They are defined as follows,

\begin{equation}\label{seplen}
\tilde{x}_i=x_i+r_i %\\
\end{equation}

\noindent where $r_i$ is the separation length between these points in direction $i$. Further, let the velocity components at ${x_i}$ and $\tilde{{x_j}}$ be $u_i$ and $\tilde{u}_j$. Starting from the Navier-Stokes equations, the transport equations of the two-point correlation tensor, $R_{ij}=\langle u_i \tilde{u}_j \rangle$, can be obtained in the following form.

\begin{equation} \label{eq1}
\begin{split}
\bigg \langle \tilde{u}_j  \frac{\partial u_i}{\partial t}  +   u_i  \frac{\partial \tilde{u}_j}{\partial t} \bigg \rangle =
%%%%%%%%%%%%
 & -  \underbrace{\bigg [    U_k \big \langle \tilde{u}_j \frac{\partial u_i}{\partial x_k} \big \rangle +  \widetilde{U}_k \big \langle u_i \frac{\partial \tilde{u}_j}{\partial \tilde{x}_k} \big \rangle  \bigg ] }_\text{$R_{ij}^A$}  
%%%%%%%%%%%%
-   \underbrace{ \bigg [   \big \langle \tilde{u}_j  u_k \big \rangle \frac{\partial U_i}{\partial x_k}  + \big \langle u_i  \tilde{u}_k \big \rangle \frac{\partial \widetilde{U}_j}{\partial \tilde{x}_k} \bigg ]  }_\text{$R_{ij}^P$} \\ 
%%%%%%%%%%%%
-  &  \underbrace{ \bigg [  \big \langle \tilde{u}_j \frac{\partial u_k u_i}{\partial x_k} \big \rangle   +  \big \langle u_i  \frac{\partial \tilde{u}_k \tilde{u}_j}{\partial \tilde{x}_k} \big \rangle   \bigg ]  }_\text{$R_{ij}^T$} 
%%%%%%%%%%%%
   - \underbrace{ \frac{1}{\rho}\bigg [  \big \langle \tilde{u}_j \frac{\partial p}{\partial x_i}   \big \rangle +  \big \langle u_i \frac{\partial \tilde{p}}{\partial\tilde{x}_j}  \big \rangle \bigg  ]  }_\text{$R_{ij}^{\Pi}$}  \\
%%%%%%%%%%%%
+    &   \underbrace{ \bigg [ \nu  \big \langle \tilde{u}_j  \frac{\partial^2 u_i}{\partial x_k \partial x_k}  \big \rangle + \nu  \big \langle u_i  \frac{\partial^2 \tilde{u}_j}{\partial \tilde{x}_k \partial \tilde{x}_k} \big \rangle \bigg ] }_\text{$R_{ij}^\nu$}
\end{split}
\end{equation}

\noindent The terms on the RHS of the equation which are labelled as $R_{ij}^A$, $R_{ij}^P$, $R_{ij}^T$, $R^\Pi_{ij}$, $R^\nu_{ij}$ are advection, production, turbulent transport, pressure and viscous terms, respectively.

The transport equations need to be simplified to perform the spectral decompositions of the various terms. In this work, the correlations are computed only in the streamwise and spanwise directions. Therefore there is no separation in the wall-normal direction (eq. \ref{ass}a). Since all the flows considered here are homogeneous in the spanwise direction, the derivative of the mean velocity with respect to $x_3$ and the mean velocity in $x_3$ is zero (eq. \ref{ass}b). Finally $u_i$ is independent of $\tilde{x}_{\alpha}$ since $u_i$ is the velocity component at another location and this is valid for $\tilde{u}_i$ and $x_\alpha$, too. Thus, their corresponding derivatives are zero (eq. \ref{ass}c).

\begin{subequations}
\begin{align}
\tilde{x}_2 &=x_2 \\
\frac{\partial U_i}{\partial x_{3}}  &= U_3 = 0\\
\text{     } \frac{\partial u_i}{\partial \tilde{x}_{\alpha}} & = \frac{\partial \tilde{u}_j}{\partial x_{\alpha}} = 0; \text{     } \alpha=1,2,3
\end{align}
\label{ass}	
\end{subequations}

We need to write equation \ref{eq1} as a function of separation lengths, $r_1$ and $r_3$, to obtain the spectral distributions. Therefore, we need to subsitute the independent variables $x_i$ and $\tilde{x}_i$. As it is usually done, the second substitution variable (associated with $r_i$) is chosen as the average position of the two points

\begin{equation} 
\Gamma_\alpha = (\tilde{x}_\alpha+x_\alpha)/2; \text{     } \alpha=1,3
\end{equation}

\noindent Let $\Phi$ be any type of two-point correlation in equation 4.2. We have the following relationships

\begin{subequations}
\begin{align}
\frac{\partial \Phi}{\partial \Gamma_3} = 0 \\
\frac{\partial \Phi}{\partial \Gamma_1} \approx 0
\end{align}
\label{corass}	
\end{subequations}

\noindent because the term $\partial \Phi/\partial \Gamma_3$ is zero due to spanwise homogeneity (eq. \ref{corass}a) and the two-point correlations are assumed to vary slowly in the streamwise direction in the case of the TBLs (eq. \ref{corass}b). The latter assumption is strong in the case of the non-equilibrium APG TBL, but it is necessary to perform the spectral decompositions. Then we perform the transformation for two-point correlations as shown in equation \ref{tv}. Note that the transformation does not apply to the derivatives of mean velocity in equation 4.2 since these derivatives are assumed constant in $x_1$ and $x_3$.

\begin{subequations}
\begin{align}
\frac{\partial \Phi_{}}{\partial x_\alpha}& = \frac{\partial \Phi_{}}{\partial \Gamma_\alpha} 
\frac{\partial \Gamma_\alpha}{\partial x_\alpha} +
\frac{\partial \Phi_{}}{\partial r_\alpha} 
\frac{\partial r_\alpha}{\partial x_\alpha} = 
- \frac{\partial \Phi_{}}{\partial r_\alpha}  \\
\frac{\partial \Phi_{}}{\partial \tilde{x}_\alpha}  & = 
\frac{\partial \Phi_{}}{\partial \Gamma_\alpha} 
\frac{\partial \Gamma_\alpha}{\partial \tilde{x}_\alpha} +
\frac{\partial \Phi_{}}{\partial r_\alpha} 
\frac{\partial r_\alpha}{\partial \tilde{x}_\alpha} = 
\frac{\partial \Phi_{}}{\partial r_\alpha} 
\end{align}
\label{tv}
\end{subequations}

\noindent Since we are interested in production and pressure-strain, we will discuss only these two ters hereafter. The production term in equation \ref{eq1} can be re-written by taking advantage of equations \ref{ass}a and \ref{ass}b.

\begin{equation} R_{ij}^{P} = 
- \big \langle \tilde{u}_j  u_1 \big \rangle \frac{\partial U_i}{\partial x_1}
- \big \langle u_i  \tilde{u}_1 \big \rangle \frac{\partial \widetilde{U}_j}{\partial \tilde{x}_1} 
- \big \langle \tilde{u}_j  u_2 \big \rangle \frac{\partial U_i}{\partial x_2}  
- \big \langle u_i  \tilde{u}_2 \big \rangle \frac{\partial \widetilde{U}_j}{\partial \tilde{x}_2} 
\end{equation}

\noindent  Furthermore, by assuming the derivatives of the mean velocity with respect to $x_1$ and $x_2$ do not change in $x_1$  as given in equation \ref{derivs}, 

\begin{equation}
\begin{split}
\frac{\partial \widetilde{U}_k}{\partial \tilde{x}_1} = \frac{\partial {U}_k}{\partial x_1}, \\
\frac{\partial \widetilde{U}_k}{\partial \tilde{x}_2} = \frac{\partial {U}_k}{\partial x_2}, 
\label{derivs}
\end{split}
\end{equation}

\noindent the production term can be rewritten as follows
\begin{equation} \label{eq:prodfull}
R_{ij}^{P} = 
- \big \langle \tilde{u}_j  u_1 \big \rangle \frac{\partial U_i}{\partial x_1}
- \big \langle u_i  \tilde{u}_1 \big \rangle \frac{\partial {U}_j}{\partial {x}_1} 
- \big \langle \tilde{u}_j  u_2 \big \rangle \frac{\partial U_i}{\partial x_2}  
- \big \langle u_i  \tilde{u}_2 \big \rangle \frac{\partial {U}_j}{\partial {x}_2} .
\end{equation}
\noindent In addition, the derivative of the mean velocity in the streamwise direction is zero for channel flows due to the streamwise homogeneity and is negligible in the ZPG TBLs. Therefore, the production term is simplified for the canonical flow cases as follows,

\begin{equation} R_{ij}^{P} = 
- \big \langle \tilde{u}_j  u_2 \big \rangle \frac{\partial U_i}{\partial x_2}  
- \big \langle u_i  \tilde{u}_2 \big \rangle \frac{\partial {U}_j}{\partial {x}_2} .
\end{equation}

%\noindent Equation \ref{eq:prodfull} will be used for the APG TBL.

\noindent The pressure term in equation \ref{eq1} is divided into two terms, 
\begin{equation}
R^\Pi_{ij} = R_{ij}^d + R_{ij}^s,
\end{equation}
\noindent where $R_{ij}^d$ and $R_{ij}^s$ are pressure transport and pressure-strain, respectively. By taking advantage of the previously introduced relationships and some mathematical manipulations, the pressure-strain term is written for the wall-parallel, $R_{\alpha\alpha}^s$, and wall-normal, $R_{22}^s$, directions as follows

\begin{equation}\label{ps}
\hspace{1.3cm}\text{ } R^s_{\alpha\alpha} = \frac{1}{\rho} 
\big \langle -  {p}\frac{\partial {u}_i}{\partial r_\alpha } +
\tilde{p}\frac{\partial  \tilde{u}_j}{\partial r_\alpha} \big \rangle; \hspace{0.5cm}\alpha=1,3
\end{equation}

\begin{equation}\label{ps}
R^s_{22} = \frac{1}{\rho} \bigg (
\big \langle   \tilde{p}\frac{\partial {u}_i}{\partial x_2 } \big \rangle +\big \langle
{p}\frac{\partial  \tilde{u}_j}{\partial \tilde{x}_2} \big \rangle \bigg ).
\end{equation}

There are several ways to perform the decomposition of the pressure term, but we chose this one for two reasons. Firstly we want to be consistent within the paper, because the spectral distributions of the channel flow that we use in the paper are obtained using the same decomposition as in \cite{lee2019spectral}. The other reason is that this decomposition is consistent with many studies in literature	 \citep{mansour1988reynolds,mizuno2016spectra} and, more importantly, with the Reynolds stress transport equation as written in equation 3.1 when $r_x$ and $r_z$ tend to zero. In other words, by integrating the spectral distributions of these terms over the wavenumbers, the budget of the Reynolds stresses, which is discussed in the previous section, is obtained.

The spectral distributions of production and pressure-strain are obtained by performing the Fourier transform of each term with respect to $r_x$ and $r_z$. In the paper, 1D spectra are computed as a function of wavenumber component $k_x$ or $k_z$ and 2D spectra as a function of $k_x$ and $k_z$.  The derivation above is for the 2D spectra, but a similar derivation is performed, albeit not given here, for the 1D spectra.

We utilize temporal data to obtain the spectra in $k_x$. We invoke Taylor's frozen turbulence hypothesis to convert frequency into streamwise wavenumber

\begin{equation}\label{taylor}
k_x  = \frac{ 2 \pi f}{ U_c}
\end{equation}

\noindent where $f$ is the sampling frequency and $U_c$ is the convection velocity which 	is taken to be the local mean velocity.	

In the paper, the spectra are always plotted as pre-multiplied by the wavenumbers, and the wavenumber axes are always in logarithmic scale. For 1D spectra, the wall-normal axis is plotted in linear scale for the outer layer and logarithmic scale for the inner layer so that both layers are examined in a more detailed, clear way.

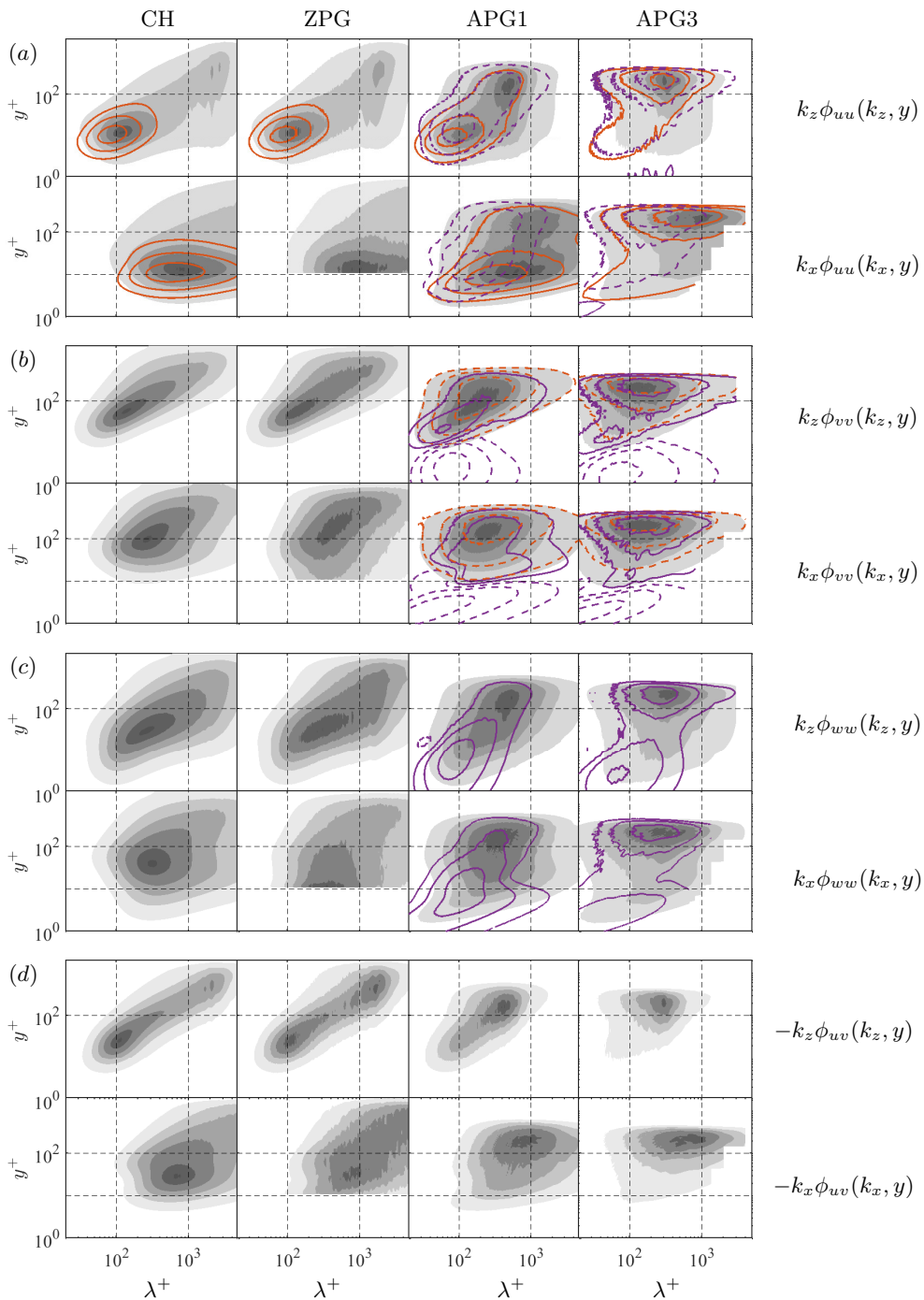
\begin{figure}
	
\vspace{0.25cm}

       \begin{tikzpicture}   
\hspace{-0.3cm}
\node(a){ \includegraphics[scale=0.45]{ 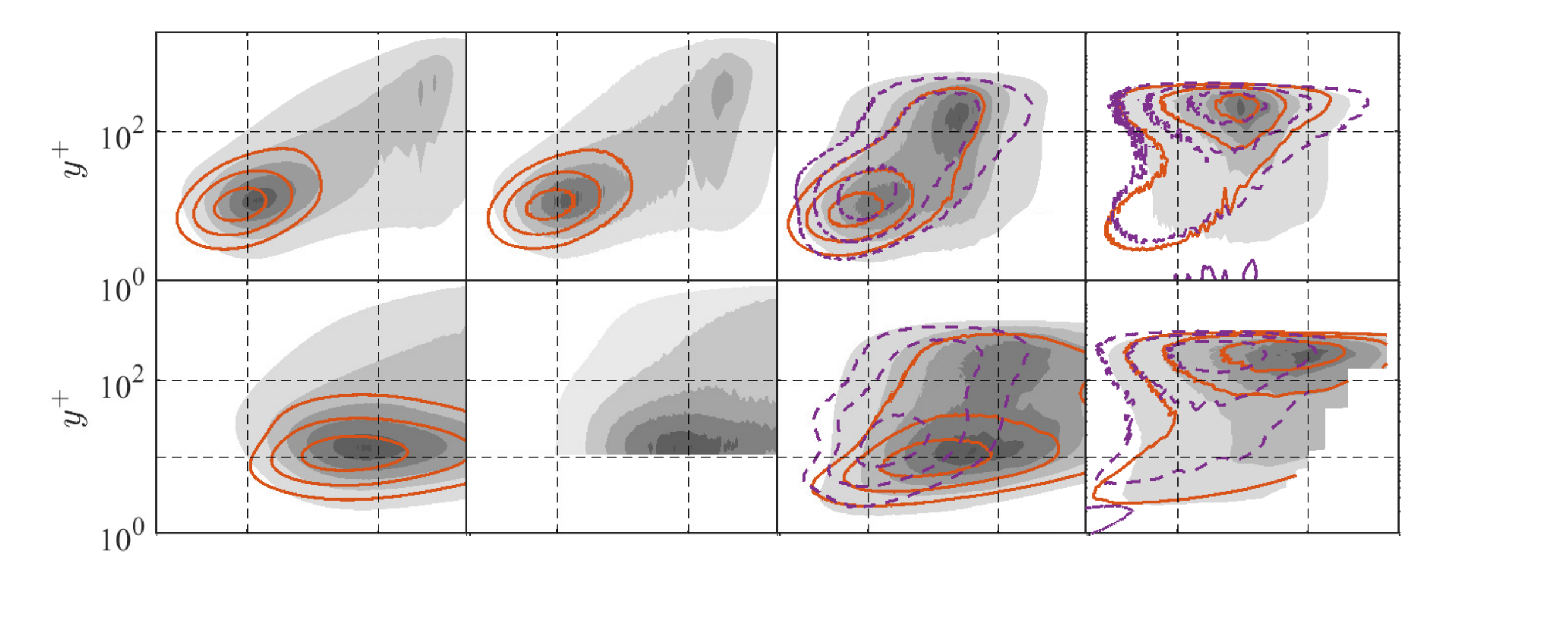}};
\node at  (-3.6,  2.5) [overlay, remember picture] {CH};
\node at  (-1.2,  2.5) [overlay, remember picture] {ZPG};
\node at  (1.2,  2.5) [overlay, remember picture] {APG1};
\node at  (3.6,  2.5) [overlay, remember picture] {APG3};
\node at  (-5.5,  2) [overlay, remember picture] {$(a)$};
\node at  (6.3,  1.2) [overlay, remember picture] {$k_z\phi_{uu}(k_z,y)$};
\node at  (6.3,-0.98) [overlay, remember picture] {$k_x\phi_{uu}(k_x,y)$};
\end{tikzpicture}

\vspace{-0.85cm}

       \begin{tikzpicture}   
\hspace{-0.3cm}
\node(a){ \includegraphics[scale=0.45]{ 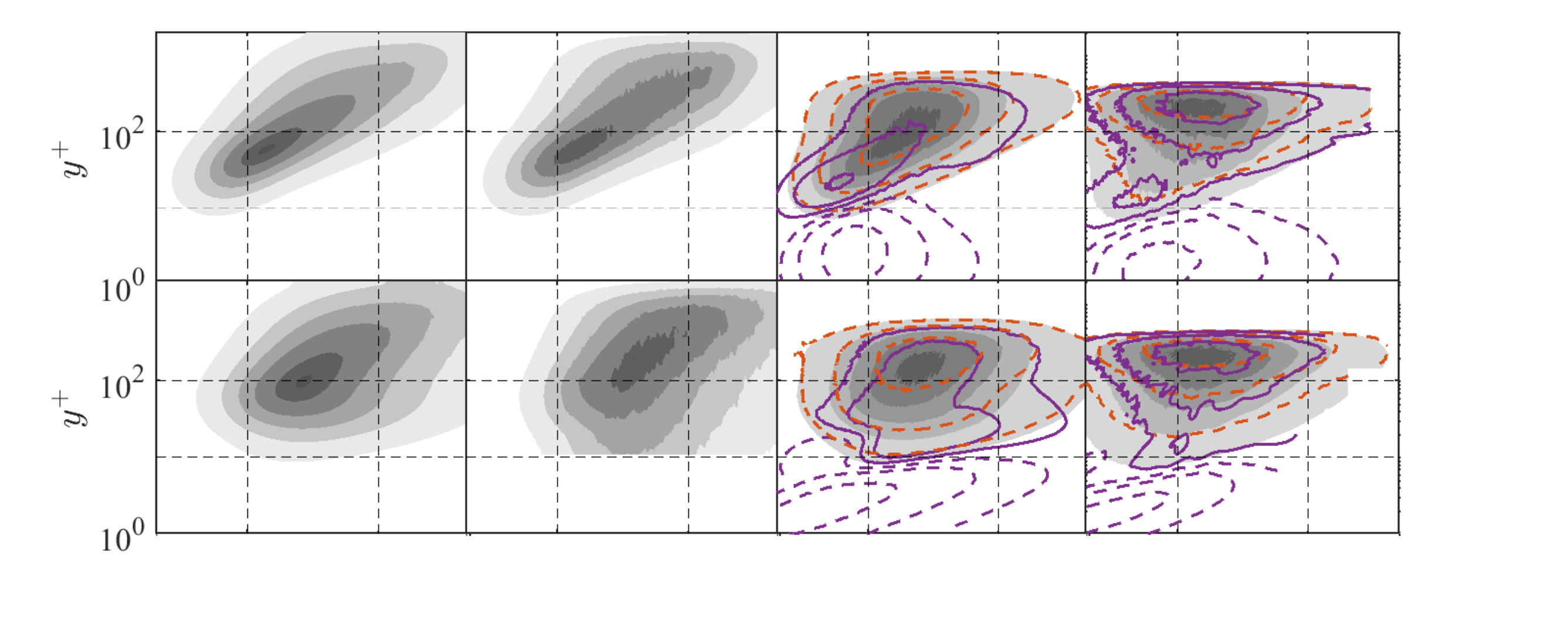}};
\node at  (-5.5,  2) [overlay, remember picture] {$(b)$};
\node at  (6.3,  1.2) [overlay, remember picture] {$k_z\phi_{vv}(k_z,y)$};
\node at  (6.3,-0.98) [overlay, remember picture] {$k_x\phi_{vv}(k_x,y)$};\end{tikzpicture}

\vspace{-0.85cm}

       \begin{tikzpicture}   
\hspace{-0.3cm}
\node(a){ \includegraphics[scale=0.45]{ 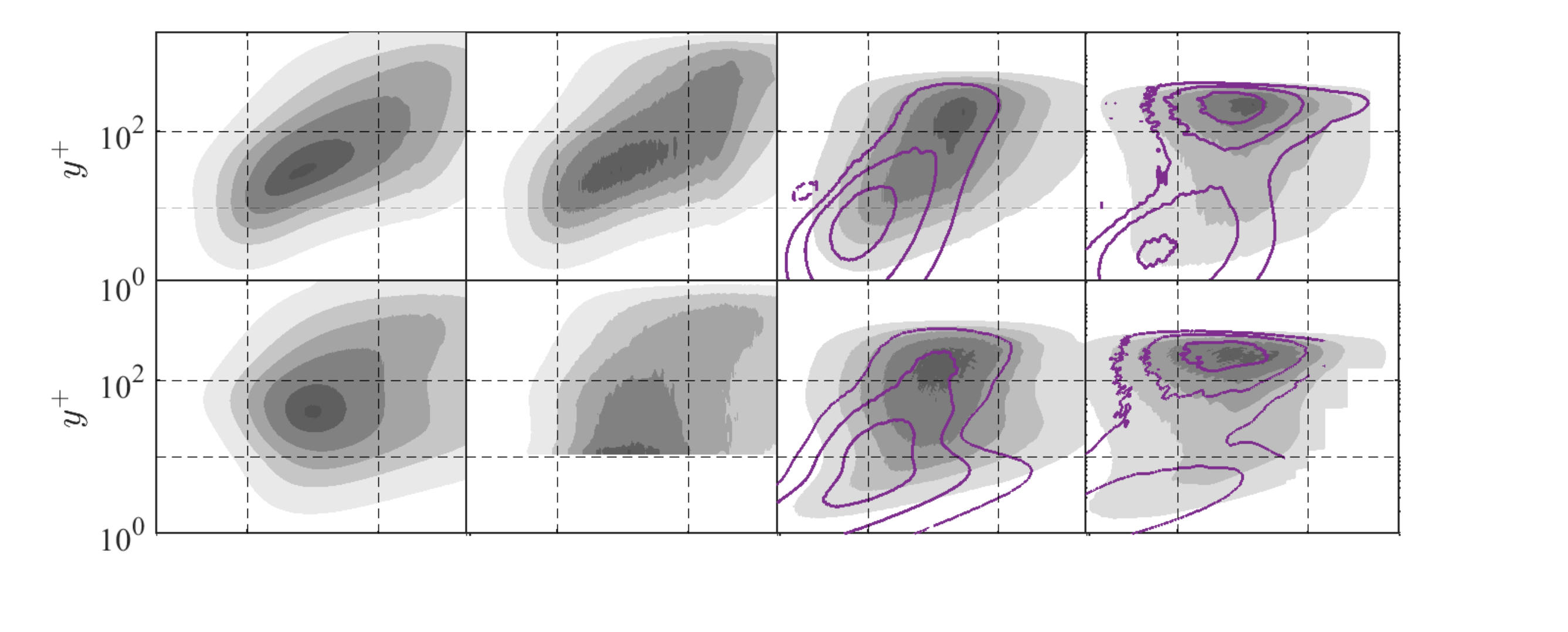}};
\node at  (-5.5,  2) [overlay, remember picture] {$(c)$};
\node at  (6.3,  1.2) [overlay, remember picture] {$k_z\phi_{ww}(k_z,y)$};
\node at  (6.3,-0.98) [overlay, remember picture] {$k_x\phi_{ww}(k_x,y)$};
\end{tikzpicture}

\vspace{-0.85cm}

       \begin{tikzpicture}   
\hspace{-0.3cm}
\node(a){ \includegraphics[scale=0.45]{ 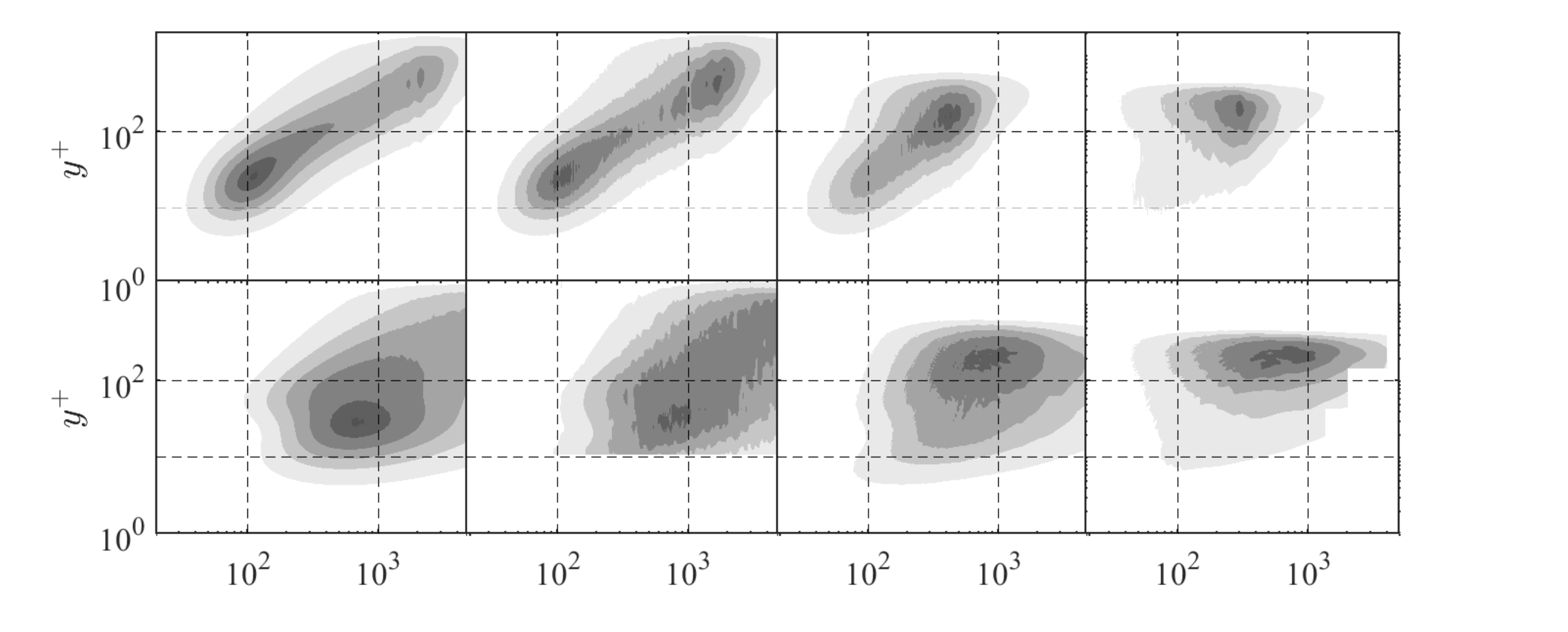}};
\node at  (-5.5,  2) [overlay, remember picture] {$(d)$};
\node at  (3.6,-2.4) [overlay, remember picture] {$\lambda^+$};
\node at  (1.4,-2.4) [overlay, remember picture] {$\lambda^+$};
\node at  (-1.2,-2.4) [overlay, remember picture] {$\lambda^+$};
\node at  (-3.6,-2.4) [overlay, remember picture] {$\lambda^+$};
\node at  (6.1,  1.2) [overlay, remember picture] {$-k_z\phi_{uv}(k_z,y)$};
\node at  (6.1,-0.98) [overlay, remember picture] {$-k_x\phi_{uv}(k_x,y)$};
\end{tikzpicture}

\caption{ The premultiplied 1D energy (shaded), production (red) and pressure-strain (purple) spectra of the channel flow (first column), the ZPG TBLs (second column), APG1 (third column), and APG3 (fourth column) as a function of $\lambda^+$ and $y^+$ for $\langle u^2\rangle$ (a), $\langle v^2\rangle$ (b), $\langle w^2\rangle$ (c) and $\langle uv\rangle$ (d). For each row the top and bottom subrows are the spanwise spectra ($k_z\phi(\lambda_z,y)$) and streamwise spectra ($k_x\phi(\lambda_x,y)$), respectively. The contour levels are $[0.1$ $0.3$ $0.5$ $0.7$ $0.9]$ of maxima of spectra for energy and $[0.1$ $0.3$ $0.7]$ of the maxima of spectra for production and pressure-strain. The dashed contours indicate negative values. The horizontal and vertical dashed lines indicate $y^+=10$, $100$ and $\lambda^+=100$ and $1000$. }
\label{fig:log_spectra}
\end{figure}

\begin{figure}
\vspace{0.25cm}

       \begin{tikzpicture}   
\hspace{-0.5cm}
\node(a){ \includegraphics[scale=0.45]{ 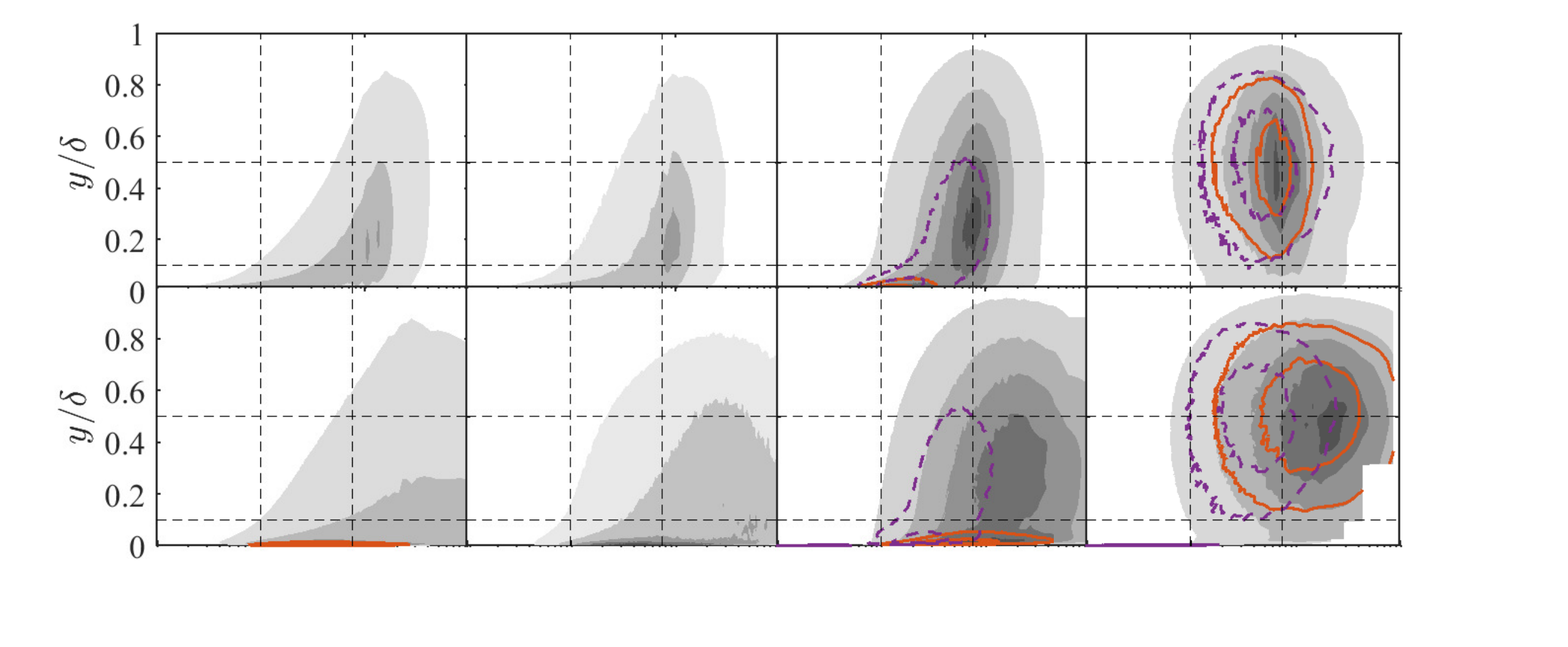}};
\node at  (-3.6,  2.5) [overlay, remember picture] {CH};
\node at  (-1.2,  2.5) [overlay, remember picture] {ZPG};
\node at  (1.2,  2.5) [overlay, remember picture] {APG1};
\node at  (3.6,  2.5) [overlay, remember picture] {APG3};
\node at  (-5.5,  2) [overlay, remember picture] {$(a)$};
\node at  (6.3,  1.2) [overlay, remember picture] {$k_z\phi_{uu}(k_z,y)$};
\node at  (6.3,-0.98) [overlay, remember picture] {$k_x\phi_{uu}(k_x,y)$};
\end{tikzpicture}

\vspace{-0.85cm}

       \begin{tikzpicture}   
\hspace{-0.5cm}
\node(a){ \includegraphics[scale=0.45]{ 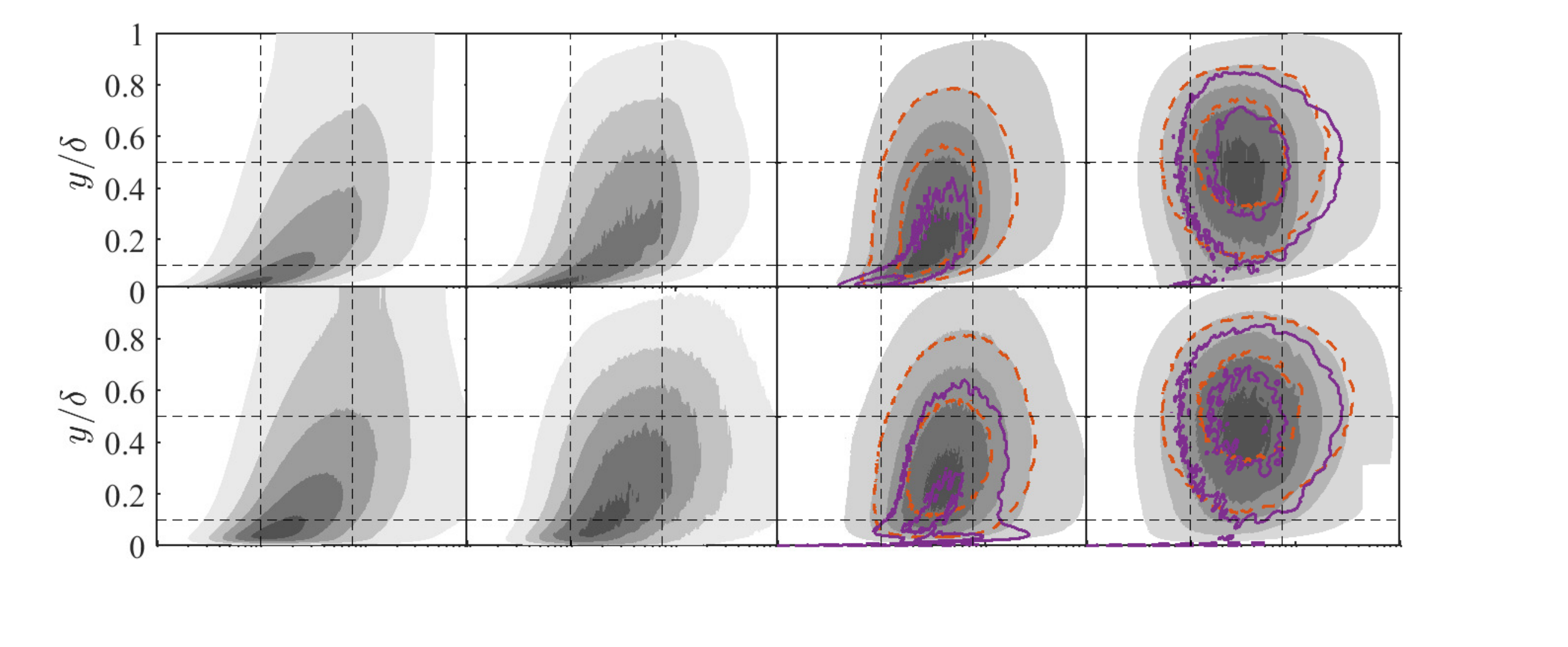}};
\node at  (-5.5,  2) [overlay, remember picture] {$(b)$};
\node at  (6.3,  1.2) [overlay, remember picture] {$k_z\phi_{vv}(k_z,y)$};
\node at  (6.3,-0.98) [overlay, remember picture] {$k_x\phi_{vv}(k_x,y)$};\end{tikzpicture}

\vspace{-0.85cm}

       \begin{tikzpicture}   
\hspace{-0.5cm}
\node(a){ \includegraphics[scale=0.45]{ 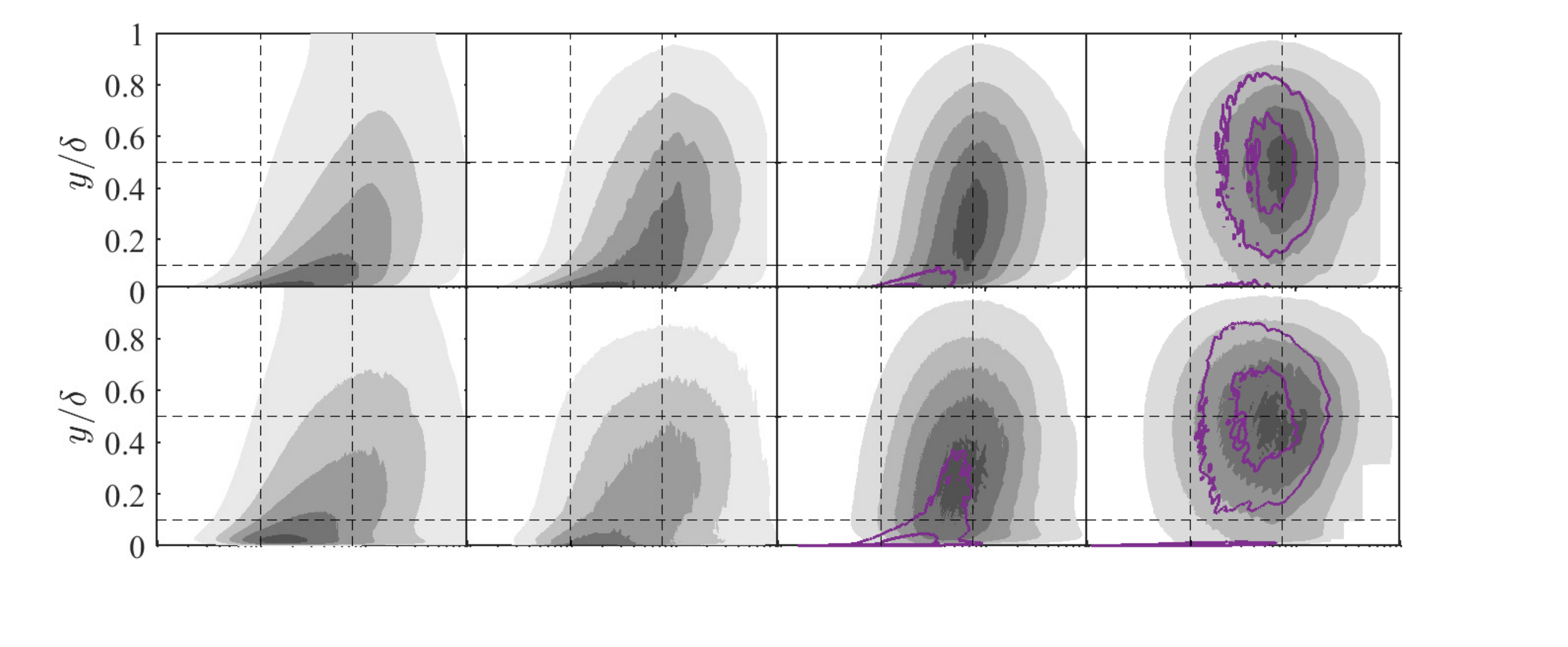}};
\node at  (-5.5,  2) [overlay, remember picture] {$(c)$};
\node at  (6.3,  1.2) [overlay, remember picture] {$k_z\phi_{ww}(k_z,y)$};
\node at  (6.3,-0.98) [overlay, remember picture] {$k_x\phi_{ww}(k_x,y)$};
\end{tikzpicture}

\vspace{-0.85cm}

       \begin{tikzpicture}   
\hspace{-0.5cm}
\node(a){ \includegraphics[scale=0.45]{ 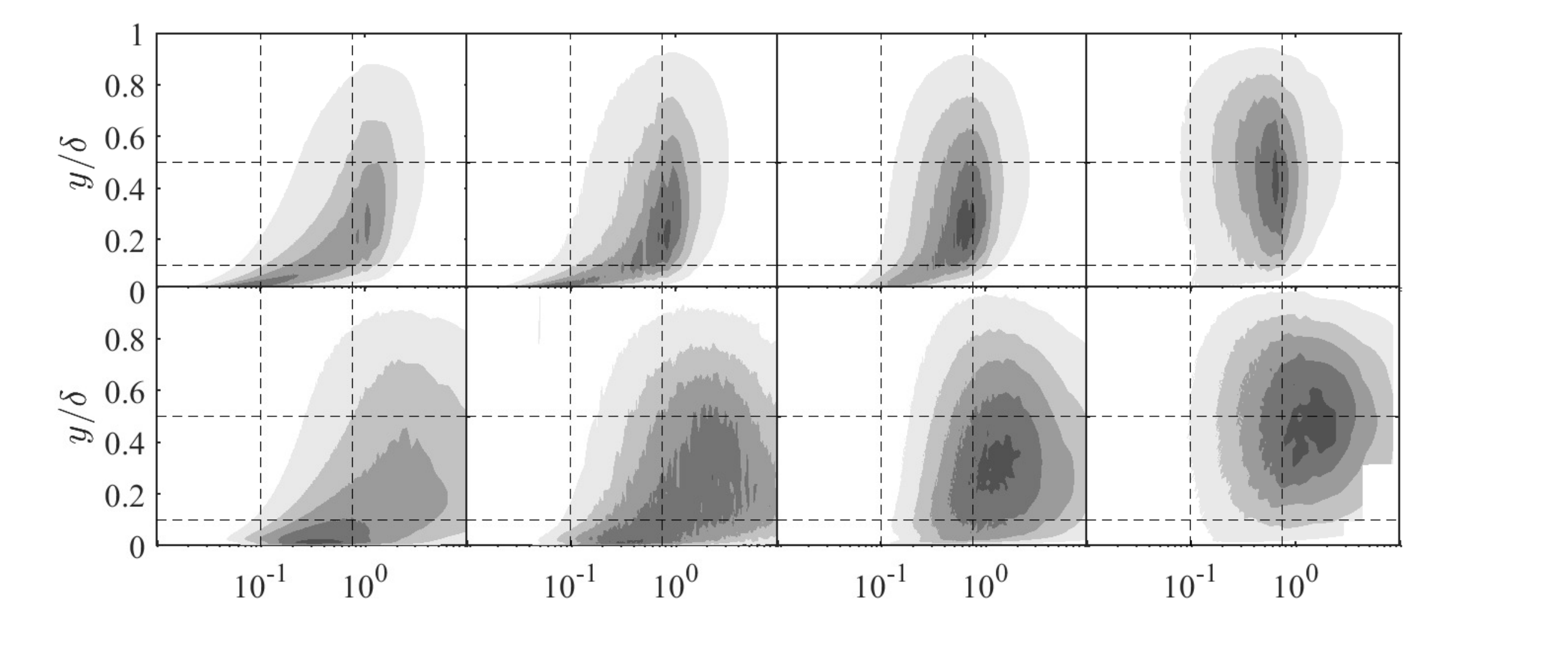}};
\node at  (-5.5,  2) [overlay, remember picture] {$(d)$};
\node at  (3.6,-2.4) [overlay, remember picture] {$\lambda/\delta$};
\node at  (1.1,-2.4) [overlay, remember picture] {$\lambda/\delta$};
\node at  (-1.2,-2.4) [overlay, remember picture] {$\lambda/\delta$};
\node at  (-3.6,-2.4) [overlay, remember picture] {$\lambda/\delta$};
\node at  (6.1,  1.2) [overlay, remember picture] {$-k_z\phi_{uv}(k_z,y)$};
\node at  (6.1,-0.98) [overlay, remember picture] {$-k_x\phi_{uv}(k_x,y)$};
\end{tikzpicture}

\caption{ The premultiplied 1D energy (shaded contours), production (red contours) and pressure-strain (purple contours) spectra of the channel flow (first column), the ZPG TBLs (second column), APG1 (third column), and APG3 (fourth column) as a function of $\lambda/\delta$ and $y/\delta$ for $\langle u^2\rangle$ (a), $\langle v^2\rangle$ (b), $\langle w^2\rangle$ (c) and $\langle uv\rangle$ (d). For each row the top and bottom subrows are the spanwise spectra ($k_z\phi(\lambda_z,y)$) and streamwise spectra ($k_x\phi(\lambda_x,y)$), respectively.  The contour levels are $[0.1$ $0.3$ $0.5$ $0.7$ $0.9]$ of maxima of spectra for energy and $[0.3$ $0.7]$ of the maxima of spectra for production and pressure-strain. The dashed contours indicate negative values. The horizontal and vertical dashed lines indicate $y/\delta=0.1$, $0.5$ and $\lambda/\delta=0.1$ and $0.75$.}
\label{fig:lin_spectra}
\end{figure}

The premultiplied 1D spectral distributions of energy (shaded), production (red) and pressure-strain (purple) as a function of $y$ and wavelength components $\lambda_x$ and $\lambda_z$ for all components of the Reynolds stress tensor are presented using inner and outer scales in figures \ref{fig:log_spectra} and \ref{fig:lin_spectra}, respectively. The 1D pressure-strain spectra are available only for the APG TBL cases. Furthermore, $\langle v^2 \rangle$ production is plotted only for the APG TBL cases due to the fact it is zero for channel flows and negligible for ZPG TBLs.

\subsection{Energy}

We start by discussing the energy spectra (shaded contours) of the channel flow and ZPG TBLs. The spectral distribution of energy shows that energy-carrying structures are mainly found in the inner layer in the channel flow and ZPG TBLs, with some activity in the outer layer. A strong peak is located at $\lambda_z^+\approx120$ and $\lambda_x^+\approx1000$ in the inner layer for the $\langle u^2 \rangle$ spectra. It is associated with the well-known streamwise streaks in the near-wall region of canonical wall-bounded flows. As for the outer layer, the  $\langle u^2 \rangle$ and $\langle uv \rangle$ distributions show an outer peak in the spanwise spectra. The streamwise spectra of CH and ZPGb do not exhibit a clear outer peak because $\Rey_\tau$ is not high enough. Channel flows and ZPG TBLs with higher $\Rey_\tau$ have a distinct outer peak in their streamwise spectra  \citep{hutchins2007large,mathis2009comparison}. This increased outer layer activity in canonical flows, as mentioned before, is attributed to the elongated meandering motions in the outer layer \citep{marusic2010high}.

The energy spectra in the small defect case of the APG TBL, APG1, demonstrate an intense turbulent activity in both inner and outer layers. The relative importance of the turbulent activity in the outer layer with respect to the inner layer is higher in APG1 than in the canonical flows. Energetic $\langle u^2 \rangle$-carrying structures are more streamwise elongated and streaky in the inner layer than in the outer layer, and the most energetic outer layer structures are slightly longer in the streamwise direction. The situation is different for the other Reynolds stress components. Even though there is energy in the inner layer, the inner peak vanishes in the streamwise and spanwise spectra. In the outer layer (figure \ref{fig:lin_spectra}), whereas $\langle uv \rangle$ carrying structures have similar streamwise lengths as $\langle u^2 \rangle$-carrying ones, the $\langle v^2 \rangle$- and $\langle w^2 \rangle$-carrying structures are shorter.

The most energetic $\langle u^2 \rangle$ structures in the inner layer of APG1 are at $y^+=15$ with $\lambda_z^+$ of $120$ and {$\lambda_x^+$ of approximately 650}, as seen from figure \ref{fig:log_spectra}. The spanwise wavelength and the wall-normal location are the same as in the channel flow and the ZPG TBL. This indicates that the wall-normal position and width of the most energetic structures are not affected by relatively small velocity defects. Similar values for $\lambda_z^+$ and $y^+$ of the inner peak have been reported in the literature for non-equilibrium \citep{tanarro2020effect} and near-equilibrium \citep{lee2017large,bobke2017history} APG TBL cases with small velocity defect, with shape factors varying from 1.42 to 1.74. Despite the similarity of the wall-normal position and $\lambda_z^+$ of the inner peak, its streamwise wavelength is smaller in APG1 than in the canonical cases where the inner peak is at $\lambda_x^+=800$ in CH and approximately at $1000$ in ZPGb. There are conflicting findings about $\lambda_x^+$ of the inner peak in APG TBLs in the literature. A reduction of $\lambda_x^+$ like in the present APG TBL has been reported by \cite{vila2020separating} in their TBL cases with shape factors varying from $1.31$ to $1.51$, as the defect increases. However, no decrease was observed by \cite{harun2013pressure} in their APG TBL case with a shape factor of $1.41$ with respect to their ZPG and FPG TBLs.

\begin{figure}
\centering 
              \begin{tikzpicture}   
       \centering           

\node(a){ \includegraphics[scale=0.75]{ 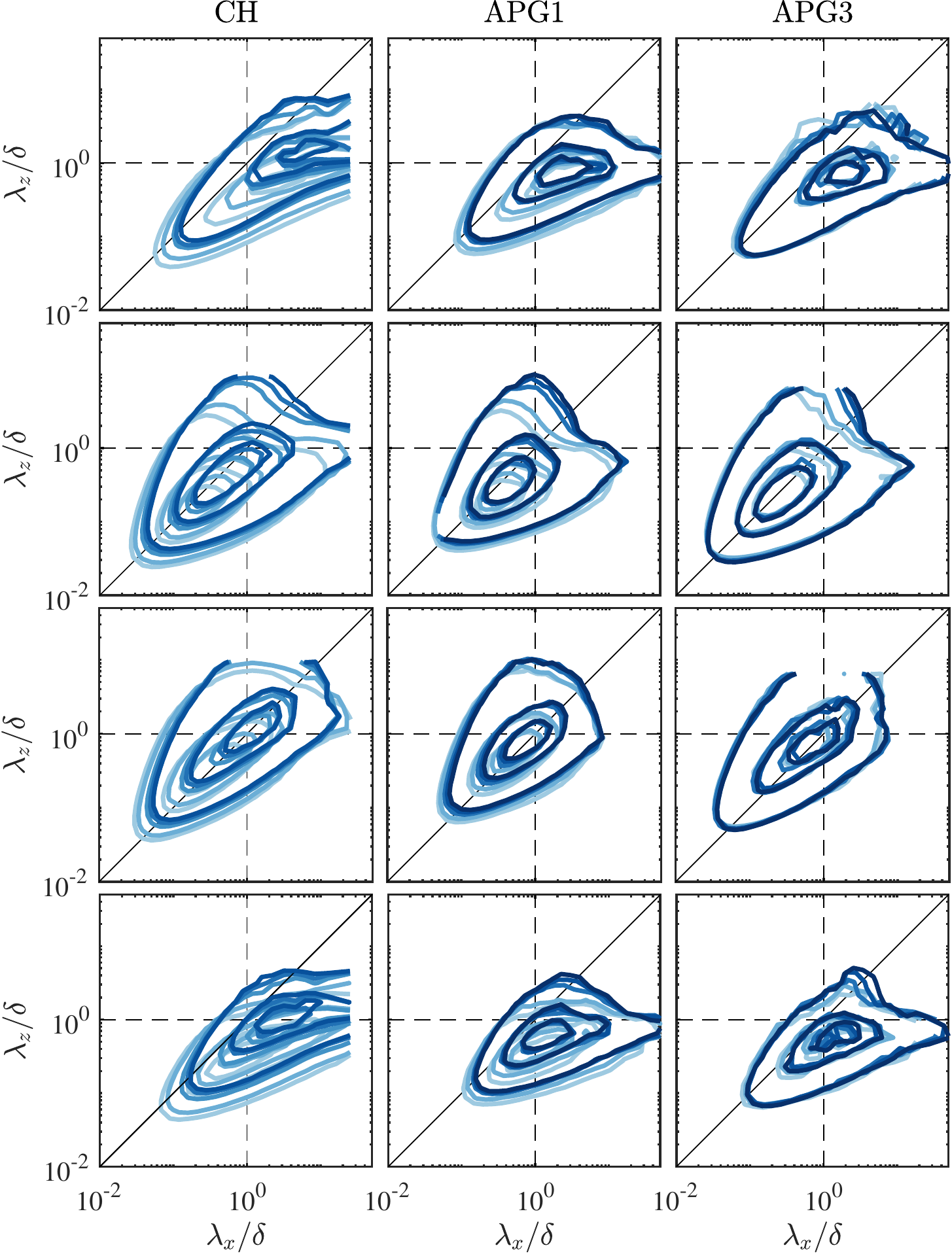}};

\end{tikzpicture}

\caption{\color{black} The premultiplied 2D energy spectra of CH (first column), APG1 (second column), and APG3 (third column) as a function of $\lambda_z/\delta$ and  $\lambda_x/\delta$ for $y/\delta=[0.1$ $0.15$ $0.3$ $0.4$ $0.5]$. The rows are for $\langle u^2\rangle$, $\langle v^2\rangle$, $\langle w^2\rangle$, and $\langle uv\rangle$. The contour levels are $[0.1$ $0.5$ $0.9]$ of the maximum of each spectra. The color is darker as $y$ increases. The straight and dashed lines indicate $\lambda_x=\lambda_z$ and $\lambda_{x,z}=\delta$, respectively.  } 
\label{rs_2d}
\end{figure}

In the large defect case, APG3, figure \ref{fig:log_spectra} shows that turbulent activity becomes much weaker in the inner layer than in the outer layer. The inner peak in the energy spectra of $\langle u^2 \rangle$ vanishes completely. Furthermore, the shape of the spectra of $\langle u^2 \rangle$ is different from the other flow cases. For instance, the energy found in the inner layer is at large wavelengths. This suggests that the large scale outer layer structures might strongly influence the inner layer in the large defect case. For the other Reynolds stress components, the trend of the inner layer losing its importance with respect to the outer layer continues as the defect increases from APG1 to APG3.

The structures carrying most of the Reynolds stresses are predominantly found in the outer layer in APG3, in the middle of the boundary layer as shown in figure \ref{fig:lin_spectra}. The situation is the same for all components. Regarding their shape, $\langle u^2 \rangle$-carrying structures are more streamwise elongated than those carrying the other components. In addition,  $\langle u^2 \rangle$-carrying ones are the longest structures with a $\lambda_x$ of $2\delta$. The most energetic structures have approximately the same $\lambda_z$, $0.75\delta $, except $\langle v^2 \rangle$-carrying ones with $\lambda_z=0.25\delta$. The prominent distinction between APG3 and APG1 is that energetic structures are located between $y/\delta=0.4-0.6$ in APG3 and $y/\delta=0.15-0.3$ in APG1. Moreover, the shape of the 1D energy spectra of $\langle u^2 \rangle$ is different between the two defect cases, which happens due to the intense inner layer activity in APG1.

To give a more complete picture of spectral properties of the energetic structures, figure \ref{rs_2d} presents the outer 2D spectral distributions of energy for CH, APG1, and APG3 as a function of $\lambda_x/\delta$ and $\lambda_z/\delta$ at several wall-normal locations (2D spectra are not available for the ZPG TBL). The $\langle u^2\rangle$ and $\langle uv\rangle$ carrying structures, especially the most energetic ones, are streaky and streamwise elongated in all flows, but they are more streamwise elongated in the channel flow than in the APG TBL cases. For instance, the peak for $\langle u^2\rangle$ spectra is at $\lambda_x\approx7$-$9\delta$ in channel flows, but it is at $\lambda_x\approx2\delta$ in the APG TBL cases.  This difference can be attributed to the presence of very-large scale structures in channel flows, which become important in high Reynolds number flows \citep{marusic2010wall}. In contrast, $\langle v^2\rangle$ and $\langle w^2\rangle$ carrying structure are not streamwise elongated. Furthermore, the shape of the most energetic $\langle v^2\rangle$ and $\langle w^2\rangle$ structures tends to follow the $\lambda_z\sim\lambda_x$ relationship in all cases even though they are smaller in the APG TBL cases. 

The energetic structures become more self-similar in the outer layer as the defect increases. In the channel flow, the smallest defect case, the shape of the 2D spectra remains similar as $y$ increases, but the size of the structures increases with respect to $\delta$ for all components. The situation is similar in APG1, but with smaller size variations. In the large defect case, the 2D spectra become self-similar starting from $y/\delta=0.1$. This trend happens probably because of increasing outer layer activity as the mean shear increases in the outer layer. As the outer layer large-scale structures start dominating the flow and the effect of the inner layer decreases, the structures that carry most of the energy become self-similar in the outer layer. 

Regarding the absolute levels of the spectral distributions (not shown in figures \ref{fig:log_spectra} to \ref{rs_2d} since relative levels are used), they increase when they are normalized with $u_\tau$ and $U_e$. This is consistent with the change of Reynolds stress profiles levels with increasing velocity defect as discussed in section 3.2.

\subsection{Production}

After discussing the energy spectra, we continue with production spectra to understand the behaviour of turbulence-producing structures. The $\langle u^2 \rangle$ and $\langle v^2 \rangle$ production are discussed separately since the former one is a source term and the latter one is a sink term that is zero in channel flows and negligible in ZPG TBLs. The inner and outer scaled $\langle u^2 \rangle$ production spectra of figures \ref{fig:log_spectra}a and \ref{fig:lin_spectra}a reflect the changes that were already observed with the production profiles of figure \ref{fig:bud_outer}. Production resides mostly in the inner layer in the case of the canonical flows, in both layers for APG1, and mostly in the outer region for APG3. The detailed characteristics of the $\langle u^2 \rangle$ production spectra in the inner and outer layers are discussed at length in Section 5.

\begin{figure}
\centering 
\begin{tikzpicture}
\node(a){ \includegraphics[scale=0.75]{ 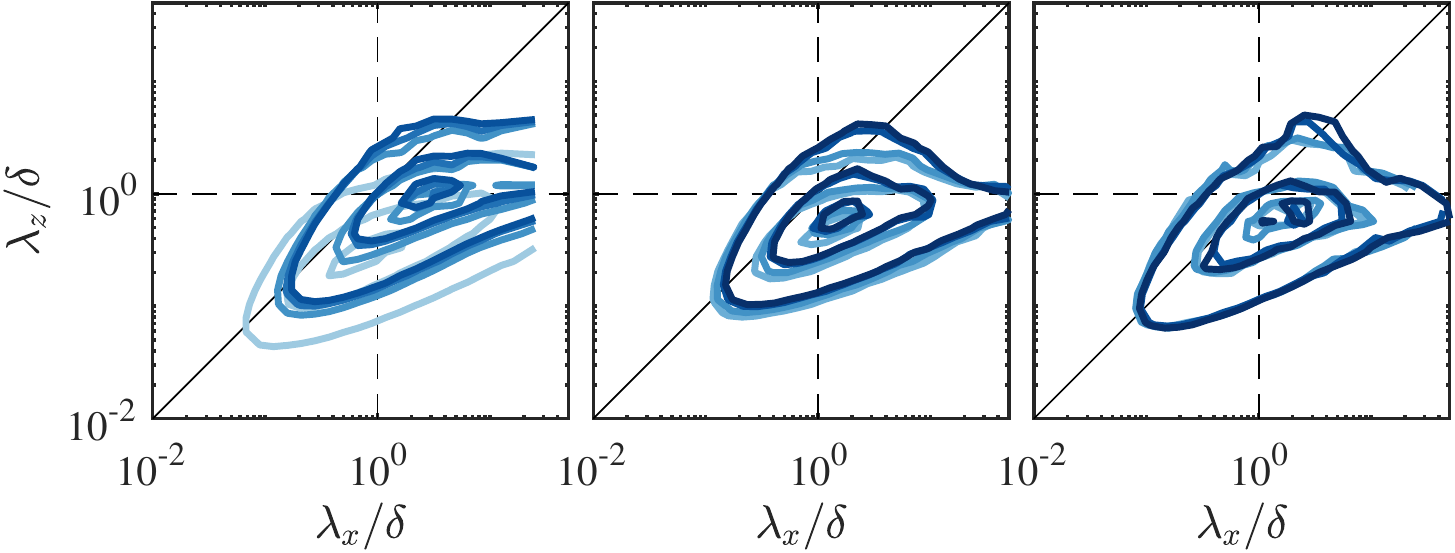}};
\end{tikzpicture}
\caption{\color{black} The premultiplied 2D production spectra of $\langle u^2\rangle$ of CH (first column), APG1 (second column), and APG3 (third column) as a function of $\lambda_z/\delta$ and  $\lambda_x/\delta$ for $y/\delta=[0.1$ $0.15$ $0.3$ $0.4$ $0.5]$. The contour levels are $[0.1$ $0.5$ $0.9]$ of the maximum of each spectra. The color is darker as $y$ increases. The straight and dashed lines indicate $\lambda_x=\lambda_z$ and $\lambda_{x,z}=\delta$, respectively.  } 
\label{prod_2d}
\end{figure}

Figure \ref{prod_2d} shows the premultiplied 2D production spectra for the $\langle u^2 \rangle$ component as a function of $\lambda_x/\delta$ and $\lambda_z/\delta$ for several wall-normal locations of CH, APG1 and APG3. Note that the $\langle uv \rangle$ spectra (figure 7, bottom row) and the $\langle u^2 \rangle$  production spectra have similar features because the spectral density function of $\langle uv \rangle$ is dominant in the production spectrum. This can be deduced from the two-point-correlation production term of equation 4.9 with $i=j=1$. The resemblance is especially strong in the case of the channel flow where production involving $\partial U/\partial x$ is zero (equation 4.10). The production structures are streaky and streamwise elongated in all cases. The shape of the spectra is similar in all cases, but with smaller structures for the APG TBLs. Whereas the streamwise-spanwise characteristics change with $y$ in the channel flow and APG1 to a lesser extent, they remain the same in the large defect case as happens for the energy spectra of figure 7.

Regarding $\langle v^2 \rangle$ production, as it was seen from the profiles in figure  \ref{fig:bud_outer}, it is zero in channel flow, negligible in the ZPG TBL cases and negative in the APG TBL cases. It is found in the outer layer of both APG TBL cases and it becomes more important with increasing velocity defect, as was mentioned previously. The spectral distributions of figure \ref{fig:log_spectra}b show that the $\langle v^2 \rangle$ production spectra are very similar to the $\langle v^2 \rangle$ energy spectra.

\begin{figure}
\centering 
              \begin{tikzpicture}   
       \centering           

\node(a){ \includegraphics[scale=0.75]{ 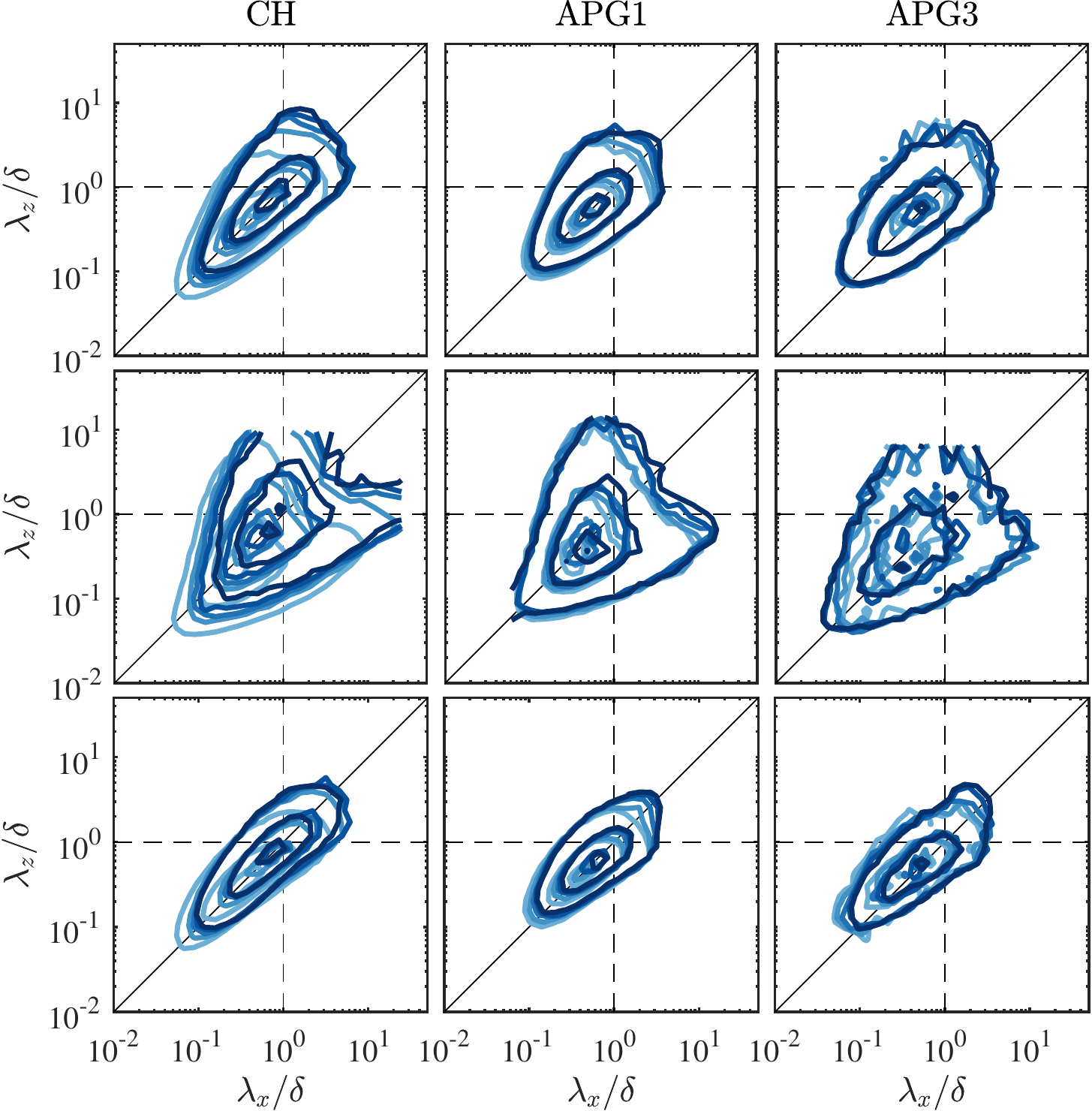}};
\end{tikzpicture}

\caption{\color{black} The premultiplied 2D pressure-strain spectra of CH (first column), APG1 (second column), and APG3 (third column) as a function of $\lambda_z/\delta$ and  $\lambda_x/\delta$ for $y/\delta=[0.1$ $0.15$ $0.3$ $0.4$ $0.5]$. The rows are $\langle u^2\rangle$, $\langle v^2\rangle$, and $\langle w^2\rangle$. The contour levels are $[0.1$ $0.5$ $0.9]$ of the maximum of each spectra. The color is darker as $y$ increases. The straight and dashed lines indicate $\lambda_x=\lambda_z$ and $\lambda_{x,z}=\delta$, respectively.  } 
\label{ps_2d}
\end{figure}

\subsection{Inter-component energy transfer}

After analyzing the turbulence-producing structures, the pressure strain spectra, as shown in figures \ref{fig:log_spectra} and \ref{fig:lin_spectra} with purple contours, are analyzed to understand the characteristics of structures that are active in inter-component energy transfer. Note that 1D pressure strain spectra are not available for the canonical flows. However, 2D pressure strain spectra are accessible for the channel flow, and are discussed further below. As previously discussed in section 3, pressure-strain governs the inter-component energy transfer. Since turbulence is transferred mostly from $\langle u^2 \rangle$ to $\langle v^2 \rangle$ and $\langle w^2 \rangle$, the $\langle u^2\rangle$ component of the pressure strain is negative, and the others are positive throughout the boundary layer, excluding the very near-wall region where the $\langle v^2 \rangle$ component is negative for both defect cases.

In the inner layer, as figure \ref{fig:log_spectra} presents, most of the inter-component energy transfer takes place with small-scale structures. In the small defect case, $\langle u^2\rangle$ transfers energy from structures wider than those that receive it for $\langle v^2\rangle$ and that are above those that receive it for $\langle w^2\rangle$. The pressure-strain structures' length are shorter than most energetic structures and the structures are less streaky. The pressure-strain spectra are qualitatively similar in the small defect case of the APG TBL and the channel flow of \cite{lee2019spectral}. In the large defect case, there is still some pressure strain in the inner layer, as happens with production, even though most of the inter-component energy transfer takes place in the outer layer.

Indeed, the inter-component energy transfer becomes dominant in the outer layer and is driven by large scale structures as the mean shear increases in the outer layer, as can be seen in figure \ref{fig:lin_spectra}. This is consistent with the behaviour of energetic and production structures. The energy is transferred from $\langle u^2\rangle$ to $\langle v^2\rangle$ and $\langle w^2\rangle$ at similar wall-normal positions for all components, unlike what happens in the inner layer. Whereas the spanwise wavelength of energy transferring outer layer $\langle u^2\rangle$ and $\langle w^2\rangle$ structures are the same, $\langle v^2\rangle$ structures are slightly narrower. The spanwise wavelengths and wall-normal positions where most of the energy transfer takes place are like those of $\langle u^2 \rangle$ production. Therefore, the outer turbulent energy is both produced and transferred to other components by structures of similar spanwise sizes and at similar locations.  In the case of the streamwise length of the dominant pressure-strain structures, it is shorter than that of 	energetic structures for $\langle u^2 \rangle$ and almost the same for $\langle v^2 \rangle$ and $\langle w^2 \rangle$.

In the outer region, narrow structures ($\lambda_z<0.3\delta$ in the small defect case and $\lambda_z<0.15\delta$ in the large defect case) transfer energy predominantly from  $\langle u^2\rangle$ to $\langle v^2\rangle$. In contrast, there is no such distinction between $\langle v^2\rangle$ and $\langle w^2\rangle$ for wide structures. Both $\langle v^2 \rangle$ and $\langle w^2 \rangle$ receive a similar amount of energy, although the levels are not given in figure \ref{fig:lin_spectra}. That energy transfer at small-scale is from $\langle u^2\rangle$ to $\langle v^2\rangle$ is also consistent with the fact that the energetic $\langle v^2 \rangle$ structures have shorter spanwise wavelengths than the energetic $\langle u^2 \rangle$ and $\langle w^2 \rangle$ structures, as discussed before.

Figure \ref{ps_2d} presents the 2D spectral distribution of pressure-strain for channel flow, APG1 and APG3 as a function of $\lambda_x/\delta$ and $\lambda_z/\delta$ for several wall-normal positions in the outer region. The structures' size notably increases in the channel flow as $y$ increases. It increases for APG1 too, but the increase is much milder. In contrast, the spectra are self-similar for APG3 throughout the outer layer in agreement with the energy and production spectra. The shape of the spectra remains qualitatively similar for each component regardless of the velocity defect. This shows that the spectral features of the pressure-strain structures are not significantly affected by the velocity defect, but their width and length slightly decrease with increasing velocity defect. Pressure-strain structures follow $\lambda_x\sim\lambda_z$.

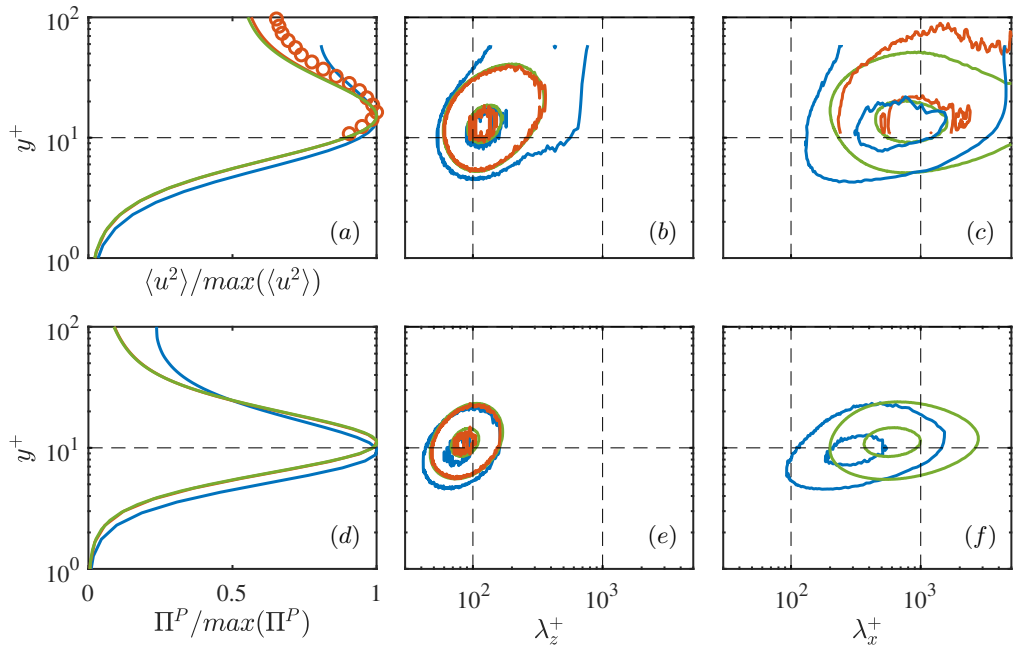
\begin{figure}
\centering 
\begin{tikzpicture}
\node(a){ \includegraphics[scale=0.75]{ 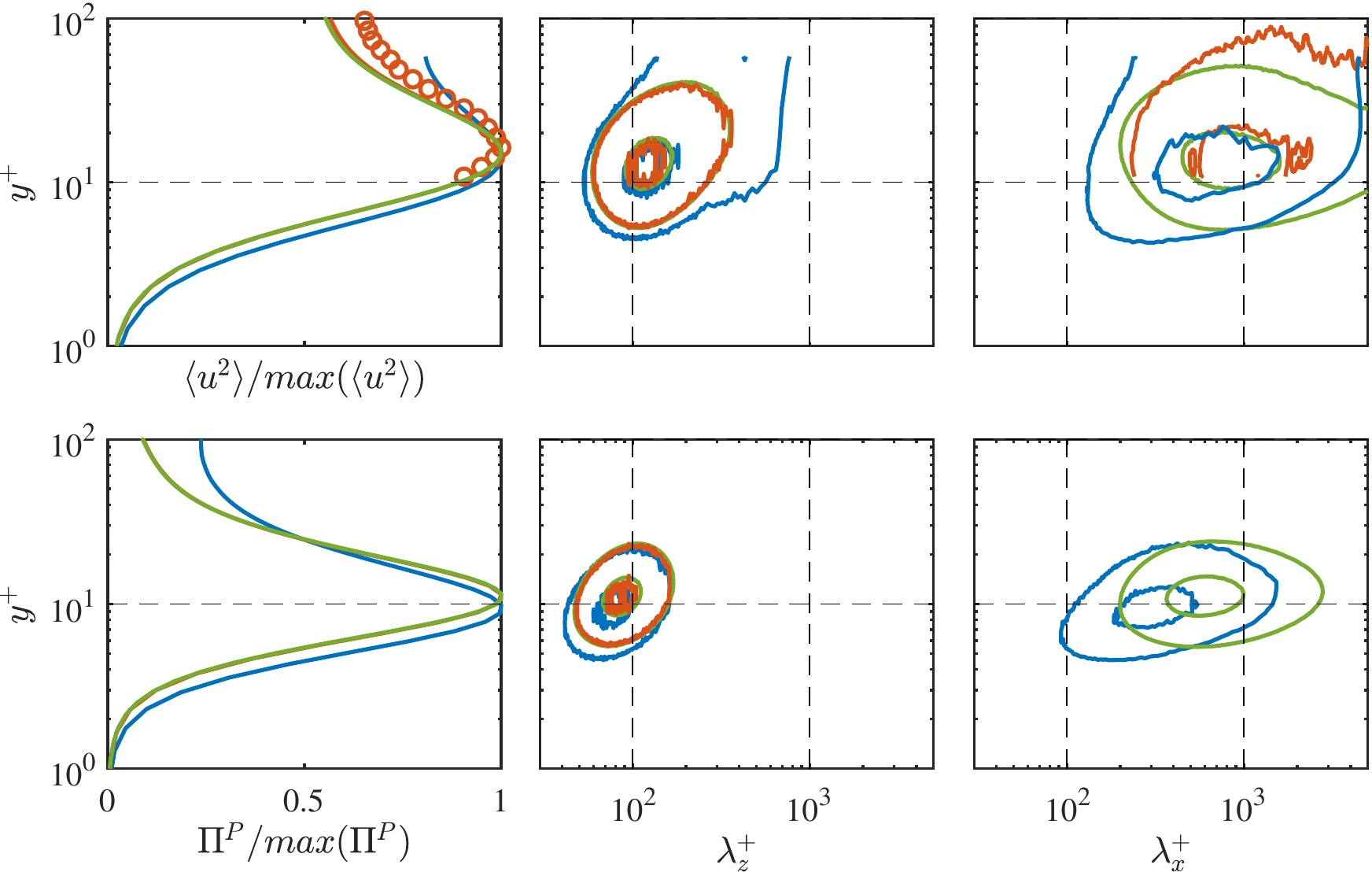}};
\node at  (-2.2,1.2) [overlay, remember picture] { $(a)$};
\node at  (-2.2,-2.8) [overlay, remember picture] { $(d)$};
\node at  (2,-2.8) [overlay, remember picture] { $(e)$};
\node at  (6.2,-2.8) [overlay, remember picture] { $(f)$};
\node at  (2,1.2) [overlay, remember picture] { $(b)$};
\node at  (6.2,1.2) [overlay, remember picture] { $(c)$};
       \end{tikzpicture}  
      
\caption{ The $\langle u^2 \rangle$ ($a$) and $\langle u^2 \rangle$ production ($d$) profiles in the inner layer as a function of $y$. The levels are normalized with the maximum of each profile. The premultiplied energy ($b$,$c$) and production ($e$,$f$) spectra of $\langle u^2 \rangle$ as a function of $y^+$ and $\lambda_z^+$ ($b$,$e$) and $\lambda_x^+$ ($c$,$f$) for APG1, the ZPG TBLs and the channel flow. The contour levels are $[0.5$ $0.9]$ of the maximum of each spectra. Colors and symbol as in Table 1.}
\label{insca}
\end{figure}

%%%%%%%%%%%%%%%%%%%%%%%%%%%%%%%%%%%%%%%%%%%%
%%%%%%%%%%%%%%%%%%%%%%%%%%%%%%%%%%%%%%%%%%%%
%%%%%%%%%%%%%%%%%%%%%%%%%%%%%%%%%%%%%%%%%%%%
%%%%%%%%%%%%%%%%%%%%%%%%%%%%%%%%%%%%%%%%%%%%
%%%%%%%%%%%%%%%%%%%%%%%%%%%%%%%%%%%%%%%%%%%%
%%%%%%%%%%%%%%%%%%%%%%%%%%%%%%%%%%%%%%%%%%%%
%%%%%%%%%%%%%%%%%%%%%%%%%%%%%%%%%%%%%%%%%%%%
%%%%%%%%%%%%%%%%%%%%%%%%%%%%%%%%%%%%%%%%%%%%
%%%%%%%%%%%%%%%%%%%%%%%%%%%%%%%%%%%%%%%%%%%%
%%%%%%%%%%%%%%%%%%%%%%%%%%%%%%%%%%%%%%%%%%%%
%%%%%%%%%%%%%%%%%%%%%%%%%%%%%%%%%%%%%%%%%%%%
%%%%%%%%%%%%%%%%%%%%%%%%%%%%%%%%%%%%%%%%%%%%
%%%%%%%%%%%%%%%%%%%%%%%%%%%%%%%%%%%%%%%%%%%%
%%%%%%%%%%%%%%%%%%%%%%%%%%%%%%%%%%%%%%%%%%%%
%%%%%%%%%%%%%%%%%%%%%%%%%%%%%%%%%%%%%%%%%%%%
%%%%%%%%%%%%%%%%%%%%%%%%%%%%%%%%%%%%%%%%%%%%
%%%%%%%%%%%%%%%%%%%%%%%%%%%%%%%%%%%%%%%%%%%%
%%%%%%%%%%%%%%%%%%%%%%%%%%%%%%%%%%%%%%%%%%%%

\section{Effects of the velocity defect on inner and outer layer turbulence}

To better understand the effects of the velocity defect on energy-carrying and energy-transferring structures, we now investigate the spectral distributions separately in the inner and outer layers because, as mentioned above, an intense activity in one layer hides what is happening in the other layer. Therefore spectral distributions are now plotted separately for each layer using their respective maxima. Furthermore, we plot spectral distributions of energy, production and pressure-strain together because we are interested in relative size and $y$-positions of structures with respect to each other. 

 Regarding the definition of the inner and outer layer, there is no commonly accepted definition for APG TBLs \citep{maciel2018outer}. For the current study, we intend to separate the inner and outer peaks when they exist. For this reason, we focus on the region $0<y^+<60$ for the inner layer and  $0.1<y/\delta<1$ for the outer region. Nevertheless, it is essential to state that these ranges may be specific to the flows studied here.

\subsection{Inner layer}

Firstly, we investigate the inner layer of the small defect case and compare it with those of the ZPG TBLs and channel flow. Figures \ref{insca}a and \ref{insca}d show the inner-scaled $\langle u^2 \rangle$ and $\langle u^2\rangle$-production profiles in the inner layer. The parameters are normalized with their maximum in the inner layer. The profiles of $\langle u^2 \rangle$ collapse perfectly for CH and ZPGa. There is a minor mismatch for ZPGb due to a decrease in levels, as explained before. The profile of APG1 is also very similar to those of the canonical flows. The peak is at $y^+=15$ in all cases. Regarding the $\langle u^2 \rangle$ production profiles (figure \ref{insca}d), the shape is very similar in all cases. The CH and ZPGa profiles collapse perfectly again, but as the defect increases, the wall-normal position of the peak location slightly decreases from $y^+=11$ in canonical flows to $y^+=9$ in APG1.

Figure \ref{insca}b, \ref{insca}c, \ref{insca}e and \ref{insca}f present the inner-scaled energy and production spectra of $\langle u^2 \rangle$ as a function of $\lambda^+$ and $y^+$ for the same flows. Like the profiles of  $\langle u^2\rangle$ and  $\langle u^2\rangle$ production, the spectra of the three flows are similar in friction-viscous units, the most noticeable difference being shorter streamwise wavelengths for APG1. The difference in streamwise wavelength is more pronounced for the production spectra (figure \ref{insca}f), the production peak is at $\lambda_x^+\approx550$ for the channel flow and $\lambda_x^+\approx300$ for APG1. For the spanwise spectra (figures \ref{insca}b and \ref{insca}e), the inner peak and contours scale well for the energy and production spectra, although the production peak is located slightly below for APG1. The aforementioned similarities in profiles and spectra suggest that the $\langle u^2\rangle$ carrying and production structures are not significantly affected by a small increase in the velocity defect, but they are definitely shortened.

\begin{figure}
\centering 
       \begin{tikzpicture}   
       \centering                     
\node(a){ \includegraphics[scale=0.75]{ 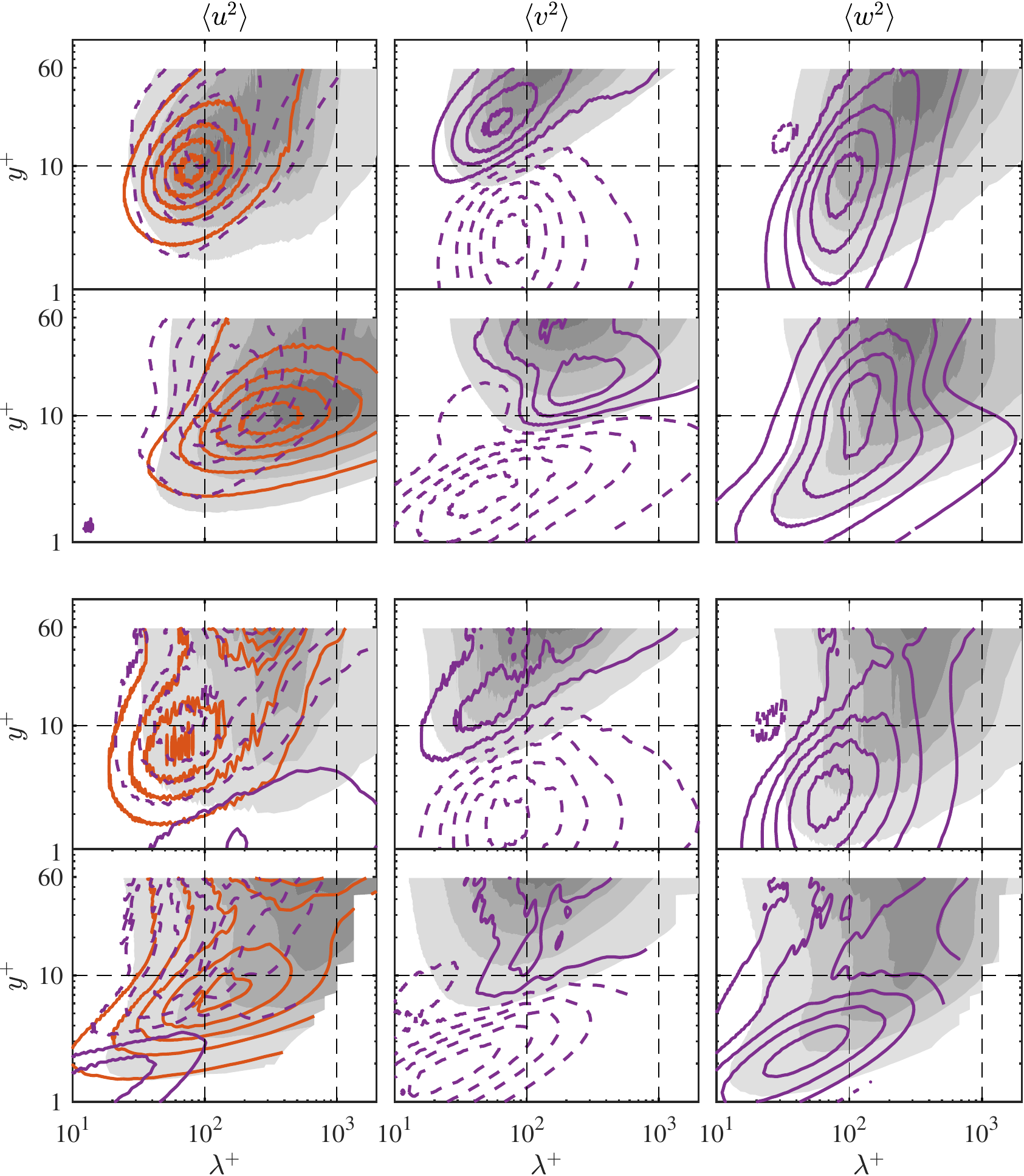}};
\node at  (-7.5,  6.6) [overlay, remember picture] {$(a)$};
\node at  (-7.5,  -0.5) [overlay, remember picture] {$(b)$};
       \end{tikzpicture}        
\caption{The premultiplied energy (shaded), production (red) and pressure-strain spectra (purple) of $\langle u^2 \rangle$ (first column), $\langle v^2 \rangle$ (second column) and $\langle w^2 \rangle$ (third column) for the inner layer of APG1 (a) and APG3 (b). For each row the top and bottom subrows are the spanwise spectra ($k_z\phi(\lambda_z,y)$) and streamwise spectra ($k_x\phi(\lambda_x,y)$), respectively. The contour levels are [0.1:0.2:0.9] of the maximum of each spectra in the inner layer. The dashed contours indicate negative values. The dashed lines denote $y^+=10$, $\lambda^+=100$ and $1000$.} 
\label{1d_inner}
\end{figure}

\begin{figure}
\centering 
              \begin{tikzpicture}   
       \centering           

\node(a){ \includegraphics[scale=0.75]{ 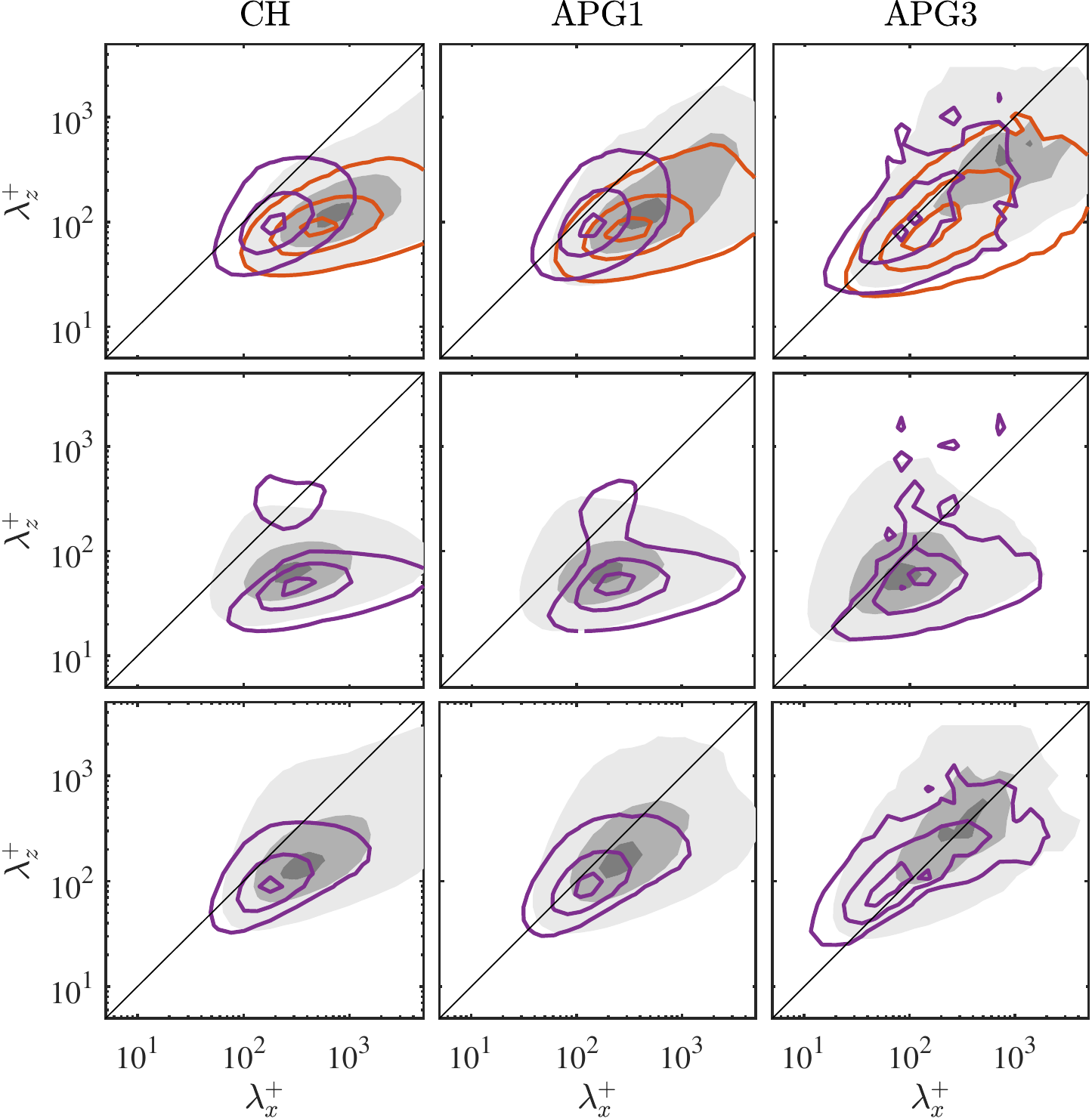}};

\end{tikzpicture}

\caption{ The premultiplied 2D energy (shaded), production (red) and pressure-strain (purple) spectra at $y^+\approx 13$. The columns are the channel flow, APG1 and APG3. The rows are $\langle u^2 \rangle$, $\langle v^2 \rangle$, and $\langle w^2 \rangle$.  The contour levels are $[0.1$ $0.5$ $0.9]$ of the maximum of each spectra.}
\label{2d_inner}
\end{figure}

After discussing the similarities between the canonical flows and the small defect case, we compare the small and large defect cases of the APG TBL in the inner layer, with this time the possibility of also comparing the pressure-strain spectra. Figure \ref{1d_inner} presents the energy, production and pressure-strain spectra for all normal components of the Reynolds stress tensor for APG1 and APG3. The outer layer is hidden, and the spectra are plotted with their maximum in the inner layer to examine the inner layer in more detail, as explained before. The most prominent difference between APG3 and APG1 is that the inner peak in the energy spectra of $\langle u^2 \rangle$ completely vanishes in the large defect case. This is rather important because the inner peak of $\langle u^2 \rangle$ is partly the signature of the streaks that are an essential part of the near-wall cycle. In contrast, both production and pressure-strain spectral distributions of $\langle u^2\rangle$ for APG3 exhibit an inner peak, although the peak is not clear because the spectra are noisy in the large defect case. As far as the shape is concerned, the shape of production and pressure-strain spectra of $\langle u^2 \rangle$ are similar in both APG TBL cases. Moreover, the relative positions and dimensions of the production and pressure-strain structures do not significantly change with increasing velocity defect. The main difference between both defect cases is the $y^+$-position and $\lambda^+$ of the inner peak of production and pressure-strain spectra of $\langle u^2 \rangle$. In APG3, the structures are closer to the wall, and their $\lambda_x^+$ and $\lambda_z^+$ are smaller than in APG1. As mentioned previously, this trend is expected because of the use of friction-viscous scales \citep{maciel2018outer}.  Nonetheless, the fact that the spectra remain similar indicates that the friction-viscous length is still relevant as a length scale for production and pressure-strain structures, at least for APG TBLs up to $H=2.6$, even if there is no significant energy in the inner layer, as shown before, and structures become irregular and disorganized \citep{maciel2017coherent}. However, the friction-viscous length cannot be a legit length scale in larger defect APG TBLs considering that the friction velocity goes to zero at separation.

To complete the picture, figure \ref{2d_inner} presents the 2D spectra of energy, production and pressure-strain for the channel flow, APG1, and APG3 as a function of $\lambda_x^+$ and $\lambda_z^+$ at $y^+\approx13$, a position close to the peaks of energy and production of $\langle u^2\rangle$. Both energetic and production structures of $\langle u^2\rangle$ are streamwise elongated, but production structures are shorter and narrower than most energetic ones for the channel flow and the small defect case. In addition, pressure-strain structures of $\langle u^2 \rangle$ are much shorter and less streamwise elongated than production and energetic structures. In the large defect case, as discussed earlier, the spectra do not show any signs of the existence of streaks that are characteristic of the near-wall cycle. Whereas the peak of the energy spectrum is at $\lambda^+_x=700$-$800$ and $\lambda_z^+=120$ for channel flow and APG1, it is at approximately $\lambda_x^+=1000$ and $\lambda_z^+=400$ for APG3. The increase in size could be due to the fact that the outer layer large-scale wide structures dominate the inner layer. This is also consistent with the 2D energy spectra further away from the wall that was discussed earlier (figure \ref{rs_2d}), the structures are much more self-similar in APG3 in the wall-normal direction. On the other hand, the production and pressure-strain structures of $\langle u^2\rangle$ are shorter in APG3, but they have similar relative sizes as in CH and APG1. Furthermore, the shape of the production and pressure-strain spectra in all cases are alike.

As for the $\langle v^2 \rangle$ component, the pressure-strain spectra have a distinctive shape in all flows. They have a second peak, weaker than the primary peak, although the peak is not clear in AGP3 due to the fact that the spectrum is noisy. This peak is at the same $\lambda_x^+$ as the primary peak, but $\lambda_z^+$ is considerably higher ($\lambda_z^+\approx300$) than in the primary peak ($\lambda_z^+\approx50$). The secondary peak might indicate that two types of pressure-strain structures play a role in receiving energy for $\langle v^2 \rangle$ in the inner layer. The primary peak's $\lambda_x^+$ decreases with increasing velocity defect from $\lambda_x^+\approx300$ to $\lambda_x^+\approx150$, while its $\lambda_z^+$ very mildly increases.

The spectra for $\langle w^2 \rangle$ are alike for CH and APG1 but are different in APG3. The shape of the spectra and the size of the structures do not considerably change as the defect increases from channel flow to APG1. The energetic structures are bigger than the pressure-strain structures. However, the energy spectrum for $\langle w^2 \rangle$ become drastically different for APG3, as happens for the $\langle u^2 \rangle$ spectrum. The $\langle w^2\rangle$-carrying structures are much broader in APG3 than in the others, but their length is similar. The peak of the energy spectra is at $\lambda_z^+\approx150$ for the channel flow and APG1 but at $\lambda_z^+\approx350$ in APG3. On the other hand, the pressure-strain structures behave in a similar fashion for $\langle v^2 \rangle$ and $\langle w^2 \rangle$. The peak of the pressure-strain spectrum is at $\lambda_z^+\approx100$ but $\lambda_x^+$ decreases from approximately $200$ to $80$ from channel flow to APG3. This behaviour is considerably different from the behaviour of the most energetic $\langle w^2\rangle$-carrying structures.

As discussed above, the velocity defect does not significantly change the spectral characteristics of the inner layer's energy-transferring structures. This suggests that the well-known near-wall cycle of the canonical flows or a turbulent regeneration mechanism with similar features to the near-wall cycle exists in APG TBLs with small and large velocity defect. Despite the similarity of energy-transferring structures, the signature of inner layer streaks, which is a crucial part of the near-wall cycle, does not exist in the $\langle u^2\rangle$ spectra of the large defect case, as shown in figures \ref{1d_inner} and \ref{2d_inner}. The lack of this signature might be explained by large-scale outer layer structures' dominating the inner layer, as mentioned above. It is possible that the footprints of these outer layer structures hide the signature of the inner layer streaks in the energy spectra. A remarkable outcome emerging from this hypothesis is that the inner layer streaks might be able to play their part in the near-wall cycle although the inner layer is dominated by much bigger structures.

\begin{figure}
\centering 
       \begin{tikzpicture}   
       \centering                     
\node(a){ \includegraphics[scale=0.75]{ 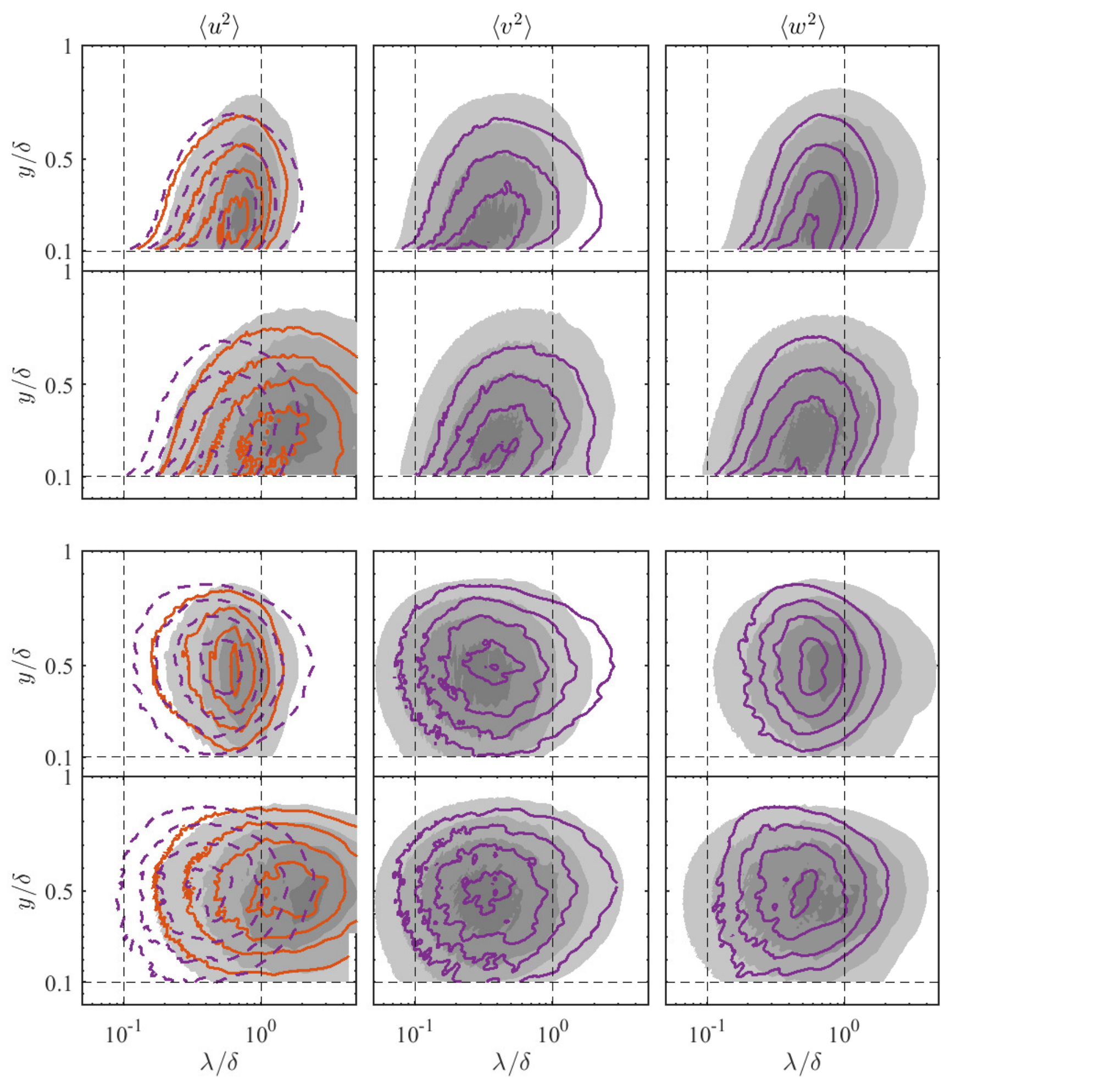}};
\node at  (-7.5,  6.6) [overlay, remember picture] {$(a)$};
\node at  (-7.5,  -0.5) [overlay, remember picture] {$(b)$};
       \end{tikzpicture}        
\caption{The premultiplied energy (shaded), production (red) and pressure-strain spectra (purple) of $\langle u^2 \rangle$ (first column), $\langle v^2 \rangle$ (second column) and $\langle w^2 \rangle$ (third column) for the outer layer of APG1 (a) and APG3 (b). For each row the top and bottom subrows are the spanwise spectra ($k_z\phi(\lambda_z,y)$) and streamwise spectra ($k_x\phi(\lambda_x,y)$), respectively. The contour levels are [0.3:0.2:0.9] of the maximum of each spectra. The dashed contours indicate negative values. The dashed lines denote $y/\delta=0.1$, $\lambda/\delta=0.1$ and $1$.} 
\label{1d_outer}
\end{figure}

\begin{figure}
\centering

              \begin{tikzpicture}   
       \centering           
\node(a){ \includegraphics[scale=0.7]{ 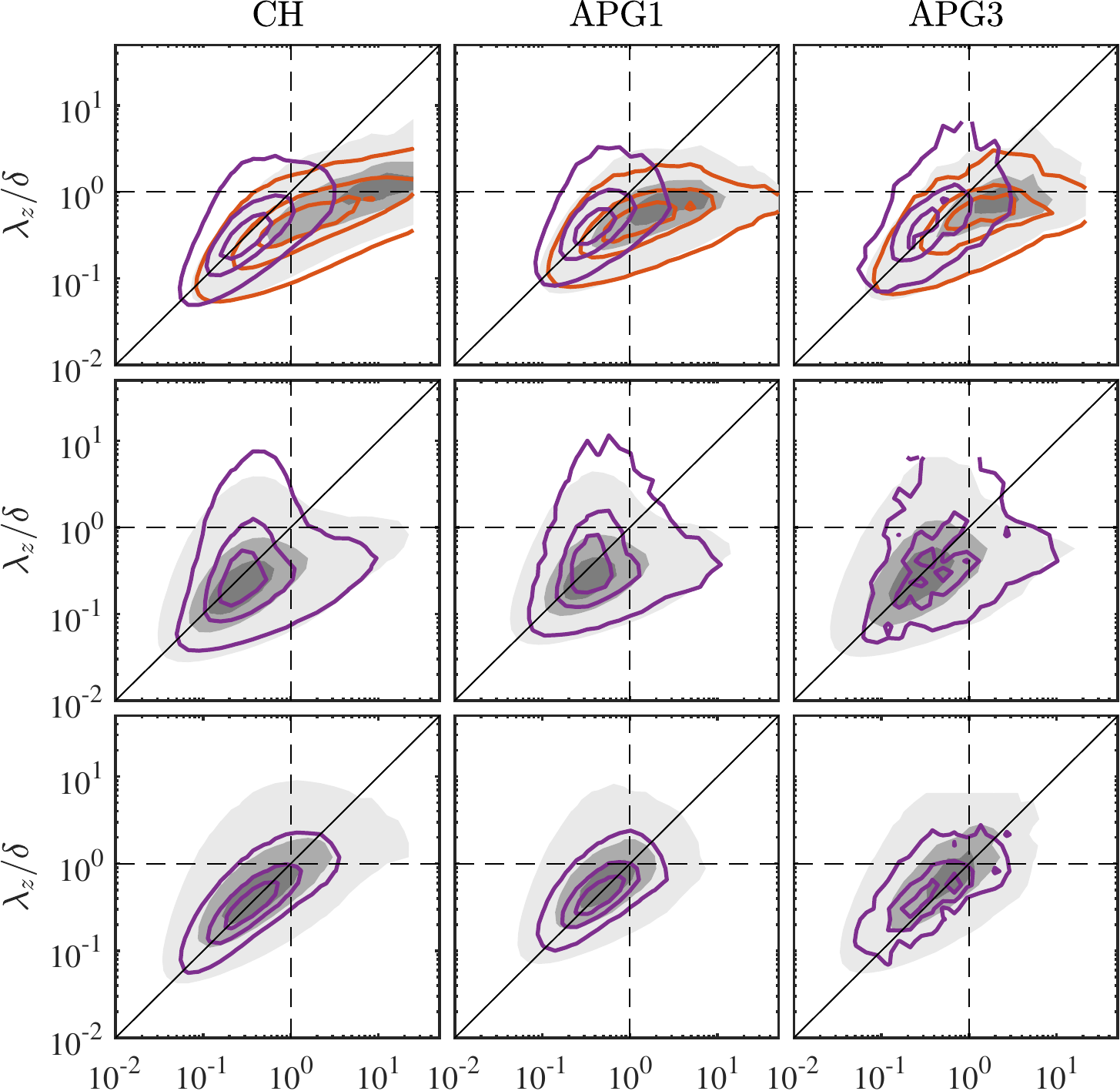}};
\node at  (-5,  4.8) [overlay, remember picture] { $a)$};

\end{tikzpicture}

      \begin{tikzpicture}   
       \centering           
\node(a){ \includegraphics[scale=0.7]{ 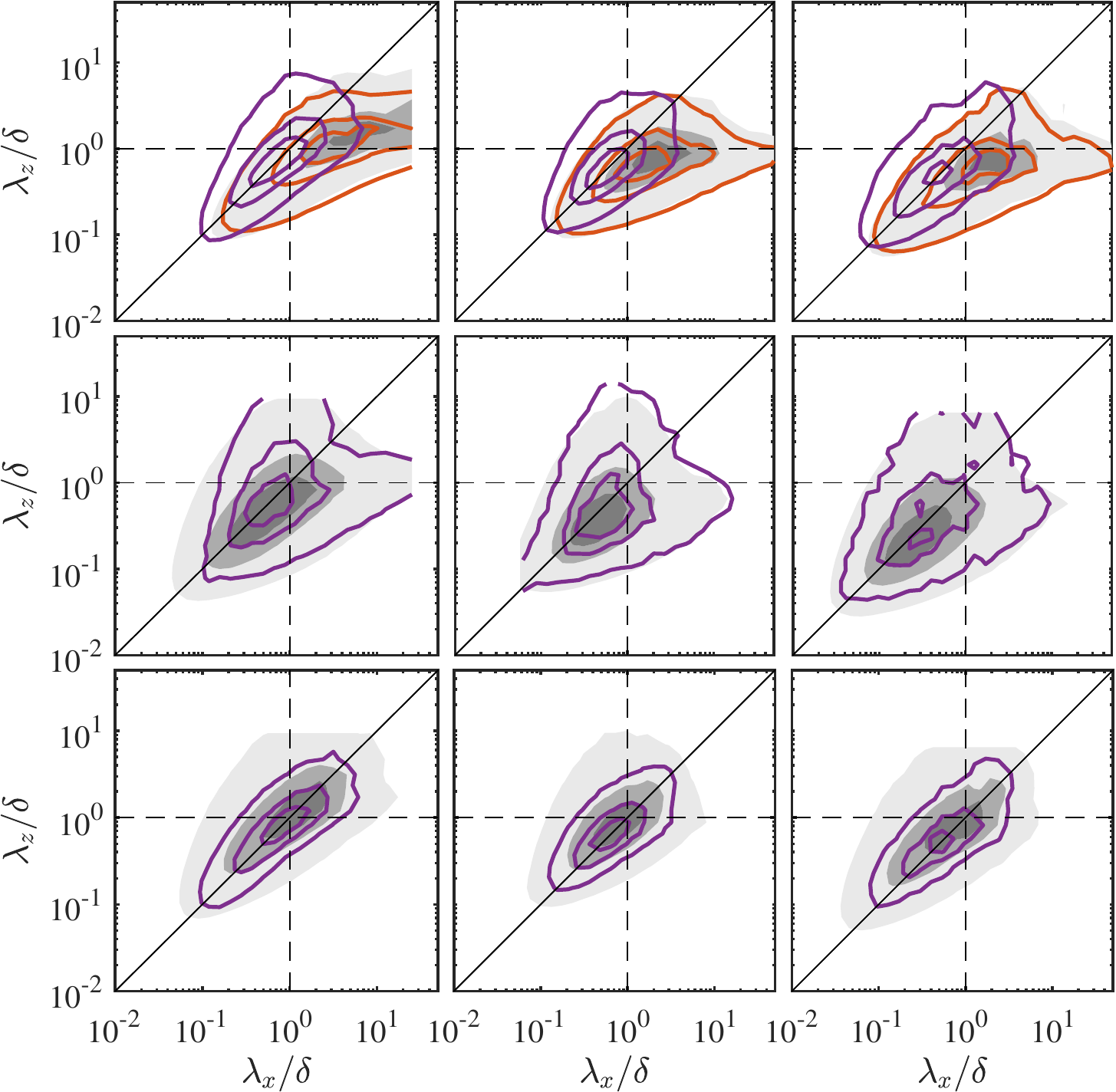}};
\node at  (-5,  4.8) [overlay, remember picture] { $b)$};

\end{tikzpicture}

\caption{ The premultiplied 2D energy (shaded), production (red) and pressure-strain (purple) spectra at $y/\delta=0.15$ ($a$) and $y/\delta=0.5$ ($b$). The columns are the channel flow, APG1 and APG3. The rows are $\langle u^2 \rangle$, $\langle v^2 \rangle$, and $\langle w^2 \rangle$.  The contour levels are $[0.1$ $0.5$ $0.8]$ of the maximum of each spectra.}
\label{2d_outer}
\end{figure} 

\subsection{Outer layer}

We present now a similar analysis for the outer layer. Figure \ref{1d_outer} displays the energy, production and pressure-strain spectra of all normal components of the Reynolds stress tensor for the small and large defect cases using a linear scale for $y$ to emphasize the outer region. Moreover, the inner layer ($y/\delta<0.1$) is masked to focus on the outer layer. Some spectral characteristics of the outer layer energy transferring structures remain similar with increasing velocity defect even though the shape of the spectra is different. For the $\langle u^2\rangle$ component, the production and pressure-strain structures are at similar $y$-locations as the most energetic structures. Whereas production structures have a similar size as the most energetic structures, pressure-strain structures are slightly narrower and considerably shorter than them in both APG TBL cases. Regarding the other Reynolds stress components, the pressure-strain structures have a similar size as the most energetic structures for $\langle v^2\rangle$ and are slightly shorter and narrower for $\langle w^2\rangle$ than the most energetic ones. Again, these relative sizes between energetic and pressure strain structures remain similar in both defect cases.

An intriguing finding about the 1D energy spectra is that the outer layer energy-carrying and -transferring structures appear to be independent of the inner layer and the wall in the large defect case. This is in contrast to the situation for the small defect case. In the latter, figures \ref{fig:log_spectra}, \ref{fig:lin_spectra} and \ref{1d_outer} show that there is considerable energy and energy transfer throughout the lower part of the boundary layer starting from the wall up to the position of the outer peaks in the energy spectra ($y/\delta\approx0.25$). Moreover, the structures' size increases with wall-normal distance in this region. This suggests the presence of wall-attached structures \citep{townsend1980structure} in the lower part of the outer layer in APG1 like in canonical flows. It was reported that the wall-attached structures carry most of the Reynolds shear stress, and hence are more dominant than wall-detached structures in canonical flows \citep{del2006self,lozano2012three}. As for the large defect case, energy and production are weak in the inner layer, and whatever activity is present appears disconnected from the strong outer layer activity. Moreover, wall distance does not scale the structures found in the lower part of the outer layer. The streamwise and spanwise wavelength of the outer layer energy-carrying and energy-transferring structures are almost the same regardless of the wall distance in the outer layer, as shown in figures \ref{rs_2d} to \ref{ps_2d}. These findings suggest that the outer layer of the large defect case is populated with detached structures. That the wall-attached structures are weaker and less numerous in large defect APG TBLs than in ZPG TBLs has been demonstrated before \citep{maciel2017coherent}. Also, it has been reported that the wall-detached structures in the outer layer are much more independent of the wall in channel flows or APG TBLs than the wall-attached ones and are similar to structures in homogeneous shear turbulence \citep{dong2017coherent,gungor2020reynolds}. Therefore, it can be concluded that the outer layer of the large defect case is dominated by structures that are not significantly affected by the wall, which is not the case in the small defect case. The dominance of wall-detached structures also suggests that the outer layer of APG TBLs could behave like free shear flows.

Despite the differences between the outer layer of the small and large defect cases discussed above, many spectral characteristics of energy-transferring structures remain similar regardless of the velocity defect. Figure \ref{2d_outer} shows the 2D spectra of the channel flow, APG1, and APG3 at $y/\delta=0.15$ and $0.50$, which are the positions with significant energy and energy transfer in the case of APG1 and APG3, respectively. The various 2D spectra have strikingly similar characteristics in all flow cases at both wall-normal positions. Their shapes, wavelength aspect ratios, and relative sizes do not change significantly with velocity defect. At $y/\delta=0.15$, the size of all structures slightly increases with increasing velocity defect, with the notable exception of $\langle u^2\rangle$ structures that are very long in the channel flow, reflecting the presence of VLSMs. The trend is reversed at $y/\delta=0.5$, all structures decrease in size with increasing velocity defect. These opposite trends possibly confirm the fact that wall-attached structures become less dominant than wall-detached ones as the defect increases. In the lower part of the outer layer, turbulent activity is strong in the small defect flows, and $y$-scaled wall-attached structures are responsible for most of this activity in canonical flows. As the defect increases, larger wall-detached structures may start to play a bigger dynamic role. This could explain, at least partly, the increase in structure size seen in the 2D spectra. Although the structure size variation goes in the other direction in the middle of the outer layer, namely a size decrease, it could still be due to the decreasing presence of wall-attached structures. Indeed, at that height wall-attached structures are big since they scale with $y$. They are bigger than wall-detached ones, contrary to the situation in the lower part of the outer layer. 
 
Even though the dominant dynamic role in the outer layer shifts from wall-attached to wall-detached structures as the defect increases, the striking spectral similarities between all flows shown in figure \ref{2d_outer} suggest that the production and inter-component energy transfer mechanisms might still be the same. This is consistent with the observation by \cite{dong2017coherent} that the fact that dynamically relevant structures are attached to the wall in channel flows is not the reason for their dynamic relevance. Wall attachment could only be a consequence of their size. Therefore, in the outer region of wall-bounded flows, the energy transfer mechanisms might remain the same no matter if dynamically relevant structures are attached or detached to the wall.

The reason for the increase of relative importance of the outer layer with respect to the inner layer might be the change in the mean shear since the spectral analysis suggests that the production mechanisms remain similar in all the considered flows. As shown previously with figures \ref{fig:mv}c and \ref{fig:mv}d, with increasing velocity defect, mean shear significantly increases in the outer layer (normalized with the outer scales) while it remains fairly constant in the inner layer (normalized with the inner scales) in comparison. Consistently with this trend, the outer layer turbulence becomes dominant as the mean shear in the outer layer increases. This trend becomes more apparent as the defect becomes large. Moreover, the present analysis shows that none of the cases examined here exhibit a peak in the $\langle v^2\rangle$ spectra with a frequency corresponding to a certain Strouhal number that would signal the presence of inflection type instabilities, although the existence of such types of instabilities in APG TBLs was reported \citep{elsberry2000experimental}. This suggests that these instability mechanisms do not importantly affect the flow or exist at all. The magnitude of the mean shear appears to be the main reason for the increase of relative importance of the outer layer activity in APG TBLs.

%%%%%%%%%%%%%%%%%%%%%%%%%%%%%%%%%%%%%%%%%%%%
%%%%%%%%%%%%%%%%%%%%%%%%%%%%%%%%%%%%%%%%%%%%
%%%%%%%%%%%%%%%%%%%%%%%%%%%%%%%%%%%%%%%%%%%%
%%%%%%%%%%%%%%%%%%%%%%%%%%%%%%%%%%%%%%%%%%%%
%%%%%%%%%%%%%%%%%%%%%%%%%%%%%%%%%%%%%%%%%%%%
%%%%%%%%%%%%%%%%%%%%%%%%%%%%%%%%%%%%%%%%%%%%
%%%%%%%%%%%%%%%%%%%%%%%%%%%%%%%%%%%%%%%%%%%%
%%%%%%%%%%%%%%%%%%%%%%%%%%%%%%%%%%%%%%%%%%%%
%%%%%%%%%%%%%%%%%%%%%%%%%%%%%%%%%%%%%%%%%%%%
%%%%%%%%%%%%%%%%%%%%%%%%%%%%%%%%%%%%%%%%%%%%
%%%%%%%%%%%%%%%%%%%%%%%%%%%%%%%%%%%%%%%%%%%%
%%%%%%%%%%%%%%%%%%%%%%%%%%%%%%%%%%%%%%%%%%%%
%%%%%%%%%%%%%%%%%%%%%%%%%%%%%%%%%%%%%%%%%%%%
%%%%%%%%%%%%%%%%%%%%%%%%%%%%%%%%%%%%%%%%%%%%
%%%%%%%%%%%%%%%%%%%%%%%%%%%%%%%%%%%%%%%%%%%%
%%%%%%%%%%%%%%%%%%%%%%%%%%%%%%%%%%%%%%%%%%%%
%%%%%%%%%%%%%%%%%%%%%%%%%%%%%%%%%%%%%%%%%%%%
%%%%%%%%%%%%%%%%%%%%%%%%%%%%%%%%%%%%%%%%%%%%

\section{Conclusions}

In the current study, we have examined the energy transfer mechanisms and energy-carrying and -transferring structures in APG TBLs and compared them with those of channel flows and ZPG TBLs. For this purpose, we have utilized the spectral distributions of energy, production, and pressure-strain of the Reynolds-stress tensor components.

The results show that spectral features of Reynolds stresses, production and pressure strain are similar in the inner layer of APG TBLs and canonical wall-bounded flows. The inner-scaled wall-normal location of the well-known inner peak of $\langle u^2 \rangle$ is the same in the small defect APG TBL case and canonical wall-bounded flows. Moreover, energetic and production structures in the inner layer have comparable spanwise wavelengths, although the streamwise length of structures is shorter in the APG TBL case than in canonical flows. When the outer layer is masked, the shape of the production and pressure-strain spectra in the inner layer is very similar in the small and large defect case, although there is no inner peak for the energy spectra of $\langle u^2\rangle$ in the large defect case. In addition, the 2D spectra of production and pressure-strain in the inner layer show similar streamwise-spanwise characteristics in the channel flow and both APG TBL cases. Therefore, energy-transferring structures and, consequently, energy transfer mechanisms in the inner layer might be the same or very similar. 

The near-wall cycle or another turbulence regeneration mechanism that works in a similar way to the near-wall cycle appears to exist in the large defect case. Interestingly, it would be present while the characteristic streaks are not the dominant structures considering their signature is absent in the energy spectra. This can be possible in two ways. The first one is that this regeneration mechanism does not need the streaks, which is rather unlikely considering the overall similarities between the flow cases. The other one is that the big outer layer structures dominate the inner layer in the large defect case in a way that the spectra do not exhibit the streak-related inner peak, but the streaks would nonetheless be present. Interestingly, this would mean that the streaks continue playing their role in turbulence regeneration even if they are amid bigger and more energetic structures.

In the outer region, the wall-normal distributions of energy, production and pressure strain spectra are different between canonical flows and APG TBLs with small and large velocity defects. The reason for this difference is that the relative intensity of the inner and outer layer turbulent activity considerably changes with increasing velocity defect, from dominant inner turbulence to dominant outer turbulence. In addition, the outer layer is dominated by wall-detached structures in the large defect case, unlike canonical flows or small defect APG TBLs where wall-attached structures are responsible for carrying most of energy and energy transfer. Due to this, the outer layer of large defect APG TBLs is much less affected by the wall than in the small defect flows. As a result, it appears to act more like a free shear layer than a wall-bounded flow. Despite these differences, the energy-transfer mechanisms might still be similar in all flow cases. The 2D spectra have strikingly similar features in the outer region of all flows, such as shape, and wavelength aspect ratios, as well as relative sizes between energy, production, and pressure-strain structures. This suggests that the distinction between wall-attached structures in small defect flows and wall-detached structures in large defect ones is not important in a dynamical sense.

 \color{black}

Strong mean shear in the outer layer seems to be the primary reason for the elevated outer layer activity in the APG TBL cases. The change of the wall-normal distribution of the Reynolds stresses, production and pressure-strain is consistent with the change of the mean shear with increasing velocity defect. Furthermore, that the spectral analysis does not show any significant difference between the energetic and energy transferring structures in channel flows and APG TBLs suggests inflection-type instabilities such as the Kelvin-Helmholtz instability either do not exist or are weak compared to other shear-driven mechanisms. Moreover, the $\langle v^2\rangle$ spectra of the large-defect APG TBL do not reveal any sign of an inflection-type instability. However, further work is necessary to confirm the presence or absence of inflection-type or other types of instabilities.

The overall conclusion is that the mechanisms producing turbulence and those responsible for inter-component energy transfer may remain the same within one layer as the velocity defect increases. What makes the large defect boundary layer different from the small defect one seems to be mainly that the mechanisms in the inner layer significantly decay in importance with respect to the outer ones with increasing velocity defect. It remains to be confirmed if this is solely due to the changes in the mean shear in each region or if other factors are at play.

\bigskip

\noindent \textbf{Acknowledgements} 

\noindent We acknowledge PRACE for awarding us access to Marconi100 at CINECA, Italy and Calcul Québec (www.calculquebec.ca) and Compute Canada (www.computecanada.ca) for awarding us access to Niagara HPC server. The authors would like to thank Myoungkyu Lee and Robert D. Moser for providing their channel flow data. TRG and AGG were supported by the research funds of Istanbul Technical University (project numbers: MGA-2019-42227 and MDK-2018-41689). TRG and YM acknowledge the support of the Natural Sciences and Engineering Research Council of Canada (NSERC), project number RGPIN-2019-04194.

%\backsection[Declaration of interests]{The authors report no conflict of interest.}

%\backsection[Author ORCID]{T.R. Gungor, https://orcid.org/0000-0002-3143-8254; Y. Maciel, https://orcid.org/0000-0003-1993-472X; A.G. Gungor, https://orcid.org/0000-0002-3501-9516}

\bibliographystyle{jfm}
\bibliography{references}

\end{document}